\documentclass[twocolumn]{aastex631}
\usepackage{apjfonts}
\usepackage{graphics,graphicx,float,import,enumitem,booktabs}

\newcommand\ii{{\sc ii}}
\newcommand\iii{{\sc iii}}

\newcommand\hii{\ion{H}{2} }
\newcommand\oiii{[\ion{O}{3}]}
\newcommand\oii{[\ion{O}{2}]}
\newcommand\siii{[\ion{S}{3}]}
\newcommand\nii{[\ion{N}{2}]}
\newcommand\sii{[\ion{S}{2}]}
\newcommand\W{$\lambda$}
\newcommand\te{T$_e$}
\newcommand\den{n$_e$}

\defcitealias{croc2006}{C06}
\defcitealias{roso2008}{R08}
\defcitealias{este2009}{E09}
\defcitealias{bres2011}{B11}
\defcitealias{tori2016}{T16}
\defcitealias{lin2017}{L17}
\defcitealias{alex2022}{A22}

\shorttitle{CHAOS VII}
\shortauthors{Rogers et al.}
\accepted{to ApJ on September 7th, 2022}

\begin{document}

\title{CHAOS VII: A Large-Scale Direct Abundance Study in M33}

\author{Noah S. J. Rogers}
\affiliation{Minnesota Institute for Astrophysics, University of Minnesota, 116 Church St. SE, Minneapolis, MN, 55455}

\author{Evan D. Skillman}
\affiliation{Minnesota Institute for Astrophysics, University of Minnesota, 116 Church St. SE, Minneapolis, MN, 55455}

\author{Richard W. Pogge}
\affiliation{Department of Astronomy, The Ohio State University, 180 West 18th Avenue, Columbus, OH, 43210}
\affiliation{Center for Cosmology \& AstroParticle Physics, The Ohio State University, 191 West Woodruff Avenue, Columbus, OH, 43210}

\author{Danielle A. Berg}
\affiliation{Department of Astronomy, University of Texas at Austin, 2515 Speedway, Austin, TX, 78712}

\author{Kevin V. Croxall}
\affiliation{Expeed Software, 100 W Old Wilson Bridge Rd Suite 216, Worthington, OH, 43085}

\author{Jordan Bartlett}
\affiliation{Department of Physics and Astronomy, University of Wyoming, 1000 E. University Ave., Laramie, WY, 82071}

\author{Karla Z. Arellano-C\'{o}rdova}
\affiliation{Department of Astronomy, University of Texas at Austin, 2515 Speedway, Austin, TX, 78712}

\author{John Moustakas}
\affiliation{Department of Physics \& Astronomy, Siena College, 515 Loudon Road, Loudonville, NY, 12211}

\begin{abstract}

The dispersion in chemical abundances provides a very strong constraint on the processes that drive the chemical enrichment of galaxies. Due to its proximity, the spiral galaxy M33 has been the focus of numerous chemical abundance surveys to study the chemical enrichment and dispersion in abundances over large spatial scales. The CHemical Abundances Of Spirals (CHAOS) project has observed $\sim$100 \hii regions in M33 with the Large Binocular Telescope (LBT), producing the largest homogeneous sample of electron temperatures (\te) and direct abundances in this galaxy. Our LBT observations produce a robust oxygen abundance gradient of $-$0.037 $\pm$ 0.007 dex/kpc and indicate a relatively small (0.043 $\pm$ 0.015 dex) intrinsic dispersion in oxygen abundance relative to this gradient. The dispersions in N/H and N/O are similarly small and the abundances of Ne, S, Cl, and Ar relative to O are consistent with the solar ratio as expected for $\alpha$-process or $\alpha$-process-dependent elements. Taken together, the ISM in M33 is chemically well-mixed and homogeneously enriched from inside-out with no evidence of significant abundance variations at a given radius in the galaxy. Our results are compared to those of the numerous studies in the literature, and we discuss possible contaminating sources that can inflate abundance dispersion measurements. Importantly, if abundances are derived from a single \te\ measurement and \te-\te\ relationships are relied on for inferring the temperature in the unmeasured ionization zone, this can lead to systematic biases which increase the measured dispersion up to 0.11 dex.

\end{abstract}

\section{Introduction}

The abundance of heavy elements in the Interstellar Medium (ISM) is entwined with the physical processes at work in a galaxy. High-mass stars forge these elements via stellar nucleosynthesis and release them into the gas via supernovae and mass-loss events. Galactic mixing processes distribute the metal-enriched gas through the ISM, while infall acts to dilute the ISM with pristine, metal-poor gas. As such, the distribution of gas-phase chemical abundances grants insight into star formation, galactic mixing mechanisms, and the overall chemical evolution of the galaxy. Observations of the ISM in spiral galaxies
typically show a negative metallicity gradient with some scatter.  That scatter, the dispersion about the abundance gradient, is a potentially powerful diagnostic of galaxy evolution.  A large dispersion indicates that enrichment processes prevail over mixing processes.  For example, if chemical enrichment is a local process, i.e., massive stars in clusters pollute their immediate environs, then large departures from the mean are possible as star formation regions at different positions in the spiral galaxy evolve independently.  On the other hand, if chemical enrichment is more of a global process, i.e., supernovae expel their newly produced heavy elements into the hot phase of the galaxy where it mixes and dilutes before cooling, then the whole galaxy experiences chemical evolution in a more homogeneous manner.

High precision measurements of the chemical abundances in a large sample of star forming regions in a individual galaxy are necessary to accurately measure the dispersion about
the mean metallicity gradient. Fortunately,
optical emission from regions of ionized gas, or \hii regions, contains a multitude of emission lines from the ions present in the gas. Despite the relatively low abundance of metal ions compared to H$^+$, emission from the forbidden transitions of O$^+$, O$^{++}$, N$^+$, and others are comparable in strength to the Balmer series. Intensity ratios of various collisionally-excited lines (CELs) from the same ion but originating from different energy levels can be exponentially sensitive to the physical conditions within the region, such as the electron gas temperature (\te) or density (\den). Provided these physical conditions, the ionic abundances of many ions can be directly calculated via the intensity of the observed CELs and the emissivities of the transitions as a function of \te\ and \den.

The direct abundance method \citep{dine1990} is often viewed as the gold standard when it comes to abundance techniques. Firstly, the emission lines necessary to measure the electron temperature and calculate ionic abundances are all within the optical, enabling ground-based observations. The optical band also contains the emission from the dominant ionization states of oxygen, which permits an accurate measure of the abundance of oxygen without the corrections for unobserved ionization states. Secondly, this abundance technique utilizes the derived physical properties within the nebula instead of relying on indicators that may not be calibrated or well-constrained at all physical conditions. Finally, this technique can be applied to multiple ionic species to obtain \te\ from numerous ionization zones within a region, thereby uncovering the temperature structure within the ionized gas.

While the direct method does have shortcomings, most notably the potential for an upward bias in temperature in the presence of temperature inhomogeneities \citep{peim1967}, presently it is the only method which allows for a robust measurement of the metallicity dispersion.  The uncertainties on strong line method metallicities of individual \hii regions are too large to accurately measure the dispersion.  Fine structure and recombination line measurements cannot provide the large sampling necessary.   

Thus, direct abundances of large, homogeneous samples are required to better understand the chemical evolution of the ISM. Local spiral galaxies present the best opportunity to study gas-phase abundances in a large quantity of \hii regions with sampling on small spatial scales. A quintessential example of such a spiral galaxy is M33; the third largest member of the Local Group, M33 is very close \citep[adopted distance of 0.86 Mpc from][]{savi2022}, relatively face-on 
(inclination $\sim$55$\arcdeg$), and hosts a number of bright \hii regions across the disk of the galaxy. Given its angular size and proximity, M33 has been the focus of many pioneering chemical abundances studies \citep{smith1975,kwitter1981,diaz1984,vilchez1988}. 
Modern, large-scale optical surveys include \hii region chemical abundances from the direct \citep{croc2006,magr2007,roso2008,magr2010,bres2011,tori2016,lin2017,alex2022} and strong-line methods \citep{lin2017,alex2022}. Additionally, there are existing recombination line \citep{este2009,tori2016} and stellar abundances \citep{U2009} in M33.

The full literature sample is one of the largest \hii region compilations in a nearby spiral galaxy, consisting of a large range in physical and ionization conditions in the ISM. However, the sample's coverage and scope is offset by its inhomogeneity. To start, a variety of detectors, all of which have varying wavelength coverage and spectral resolution, are used to obtain the optical spectra. Furthermore, the \hii regions selected within each sample are optimized for different studies of the ISM. For example, samples targeting many \hii regions across the disk of the galaxy are beneficial for an accurate measure of the abundance gradient and dispersion about it \citep[e.g.,][]{roso2008}. Other samples \citep[e.g.,][]{este2009,tori2016} focus on relatively few, bright regions for the most reliable direct and recombination line (RL) abundances for insight into the Abundance Discrepancy, or the consistent trend that RL abundances are significantly larger (around a factor of 2 in \hii regions) than CEL abundances as measured in the same object \citep{peim2005,garc2007,este2009,garc2013}. Many studies in the literature were observed at relatively low spectroscopic resolution ($\ge$ 5 \AA) which is insufficient for isolating key emission lines from other emission lines or atmospheric features. Finally, each study selects its own atomic data, reddening determination method, electron temperature relations, etc., based on the available spectroscopic data. As such, the literature temperatures and/or abundances in M33 cannot necessarily be compared on a simple one-to-one basis.

To better understand the chemical evolution of this important galaxy, and to complement previous chemical abundance studies, the CHemical Abundances Of Spirals (CHAOS) project \citep{berg2015} has observed a large population of \hii regions in M33. To date, CHAOS has accumulated 200+ high-resolution \hii region spectra in nearby, face-on spiral galaxies. With this database, CHAOS has measured statistically-significant direct oxygen abundance gradients in five galaxies, found evidence of universal secondary N/O gradients in local spirals, measured \ion{C}{2} recombination line abundances of bright regions, and developed robust empirical \te-\te\ relations and a new method of application for these relations \citep{berg2015,crox2015,crox2016,berg2020,skil2020,roge2021}. The \hii regions of M33 with auroral line detections increases the database by almost 33\%, but the real advantage is its wealth of direct abundance data and homogeneity.

This study is organized as follows: In \S2 we introduce the observations and reduction of the CHAOS M33 data, and present the literature samples we compare to. The parameters used for direct abundances, including \te\ and Ionization Correction Factors (ICFs), are discussed in \S3. The oxygen abundances, gradient, and dispersion in M33 as well as the abundance of other heavy elements (N, Ne, S, Cl, Ar) are reported in \S4. We compare our results to the literature, discuss possible sources of \te\ contamination (some unique to M33), and contextualize the abundance dispersion observed in local spiral galaxies in \S5. We summarize our conclusions in \S6.

\section{Observations and Reduction}
\vspace{-2mm}
\begin{deluxetable}{lccc}  
\tablecaption{M33 Global Properties}
\tablewidth{\columnwidth}
\tablehead{ 
  \colhead{Property}	&
  \colhead{Adopted Value}	&
  \colhead{Reference}	
  }
\startdata
R.A. 	& 01$^h$33$^m$50.6$^s$ &  1 \\  
Decl.  & +30$^\circ$39$^m$29.9$^s$ & 1 \\
Inclination & 55.08$^\circ$ & 1\\
Position Angle  & 201.12$^\circ$ & 1\\
Distance & 859$^{+24}_{-23}$ kpc & 2 \\
log(M$_\star$/M$_\odot$) & 9.68 & 3\\
R$_e$ & 555\arcsec, 2.31 kpc & 4 \\
Redshift & $-$0.000597 & 1
\enddata
\label{t:m33global}
\tablecomments{Units of right ascension are hours, minutes, and seconds, and units of declination are degrees, arcminutes, and arcseconds. References: [1]  \citet{koch2018}  [2] \citet{savi2022} [3] \citet{corb2014} [4] \citet{lero2019,lero2021}}
\vspace{-6mm}
\end{deluxetable}
\begin{figure*}[t] 
   \centering
   \includegraphics[width=0.9\textwidth, trim=30 0 30 0,  clip=yes]{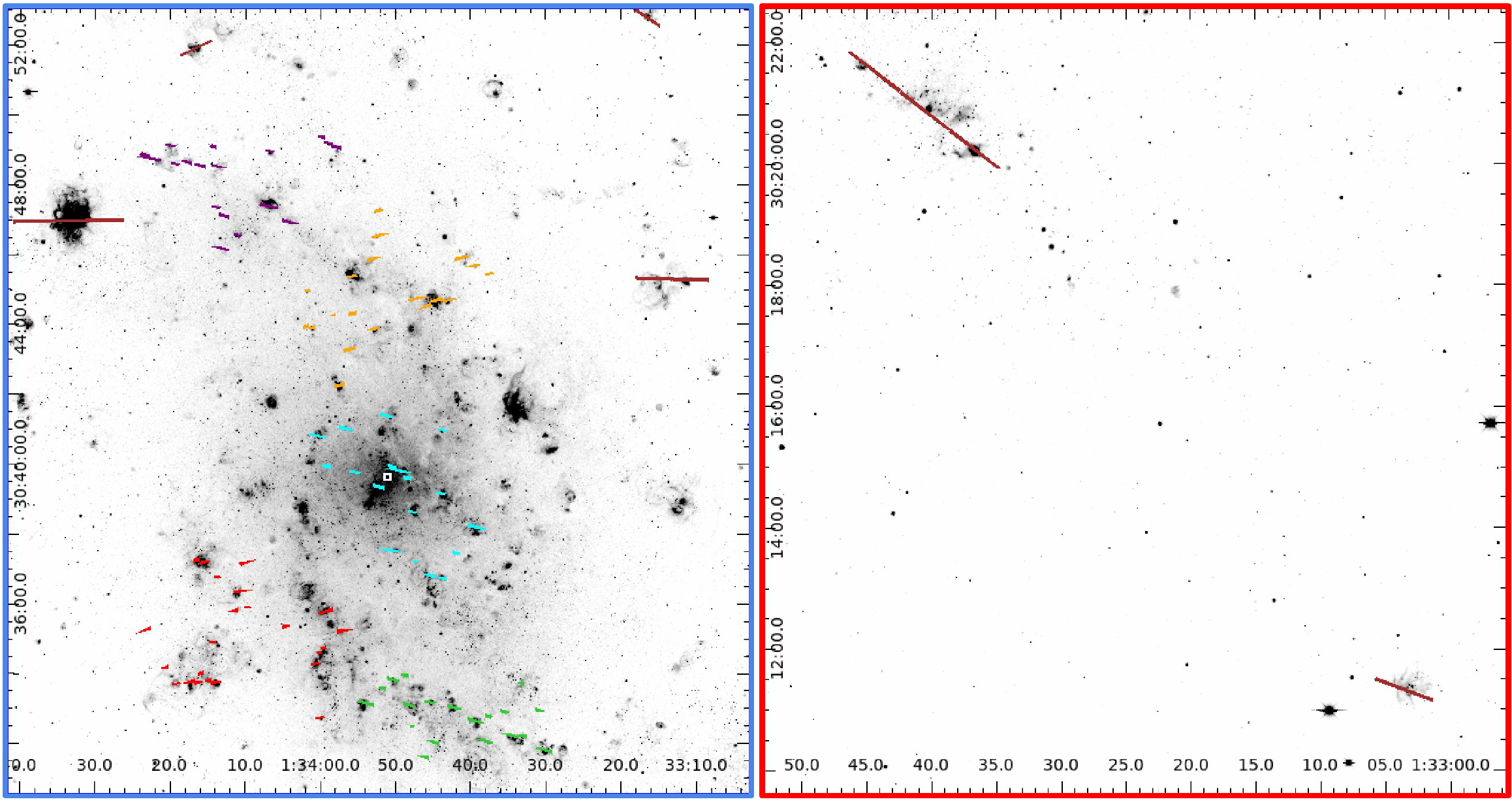}
   \caption{A continuum-subtracted H$\alpha$ image of M33 from \citet{mass2006} retrieved from \citet{https://doi.org/10.26132/ned1} with the CHAOS slits superimposed on the targeted regions. The R.A. and Decl. are reported on the vertical and horizontal axes, respectively. \textit{Left}: H$\alpha$ image of the center of M33. This image encompasses the five MOS fields (Red, Orange, Green, Cyan, and Purple slits) and four longslit pointings (Brown slits). \textit{Right}: H$\alpha$ image focused on the far southern portion of M33. Two longslit pointings are provided as the Brown slits.}
   \label{fig:halphaimg}
\end{figure*} 

\subsection{CHAOS Observations of M33}

The CHAOS project utilizes the Multi-Object Double Spectrographs \citep[MODS,][]{pogg2010} on the Large Binocular Telescope \citep[LBT,][]{hill2010} to obtain the optical spectra of \hii regions in nearby spiral galaxies. The MODS blue channel has a wavelength coverage of 3200--5700 \AA\ and R$\sim$1850 for the G400L (400 lines mm$^{-1}$) grating; the red channel has a wavelength range of 5500--10000 \AA\ and R$\sim$2300 for the G670L (250 lines mm$^{-1}$) grating. In combination, these spectrographs cover the full optical band and extend into the NIR at sufficient resolution for direct abundance analysis. Multi-object slit (MOS) masks can observe $\sim$20 objects in one 6\arcmin $\times$6\arcmin\ field of view, while longslits can target objects that are too extended (in galactocentric radius or in emission) for the MOS masks. The versatility, sensitivity, resolution, and wavelength coverage of MODS permits a thorough examination of the direct abundances in M33 and other nearby spiral galaxies.

For M33's parameters, we adopt the disc parameters of center, position angle, and inclination that \citet{koch2018} derive from a fit to the velocity field derived from combined VLA and GBT \ion{H}{1} 21 cm observations. These properties are provided in Table \ref{t:m33global}. Note that the single values of position angle and inclination do not account for the outer warp \citep[cf.,][]{corb2014} but that feature appears beyond $\sim$ 8 kpc, and is not relevant for our \hii region sample. There are many distances to M33 in the literature to choose from \citep[see discussion in][]{degr2017,lee2022}.  We adopt the distance modulus of 24.67$\pm$0.06 (corresponding to 859$^{+24}_{-23}$ kpc) based on HST observations of RR Lyrae \citep{savi2022} anchored to the GAIA eDR3 reference frame \citep{naga2021}, which is consistent with the values favored by \citet{degr2017} and \citet{lee2022}. The effective radius of M33, R$_e$, is determined from the z0MGS WISE 3.4$\mu$m maps \citep{lero2019} with the same fitting method as described in \citet{lero2021} and which has been utilized in previous CHAOS studies \citep[e.g.,][]{berg2020}.

Given its proximity and wealth of \hii regions, CHAOS observed five MOS fields in M33 with an additional six longslit pointings. All MOS observations and the majority of the longslit observations took place between 2015 October 11 and 15; the remaining longslit pointings were taken 2015 December. Each MOS field was observed in six 1200 second exposures, while longslit pointings were observed for three 1200 second exposures. The standard star G191-B2B was observed on multiple nights for flux calibration. The width of each slit in the MOS masks was 1\farcs0 and the length ranged from 6\farcs0--30\farcs0; the longslits are a combination of five 60\farcs0$\times$1\farcs0 slits. The airmasses of the observations varied depending on the time of observation, and the position angle was chosen to be equal to the parallactic angle halfway through a pointing to minimize flux losses due to differential atmospheric refraction \citep{fili1982}. In total, 99 slits target ionized regions based on their surface brightness, location, and overlap with previous direct abundance studies. Figure \ref{fig:halphaimg} provides a continuum-subtracted H$\alpha$ image of M33 along with the MOS and longslit locations of the CHAOS observations. Table \ref{t:m33Obs} provides the name, location, and radius for each \hii region observed in M33. Consistent with previous works, we report the name of the \hii region as the offset in R.A.\ and Decl.\ of the center of the extraction profile relative to the center of the galaxy (provided in Table \ref{t:m33global}).

\subsection{Data Reduction and Processing}

\begin{figure*}[!t]
\epsscale{1.0}
   \centering
   \includegraphics[width=0.7\textwidth, trim=30 0 30 0,  clip=yes]{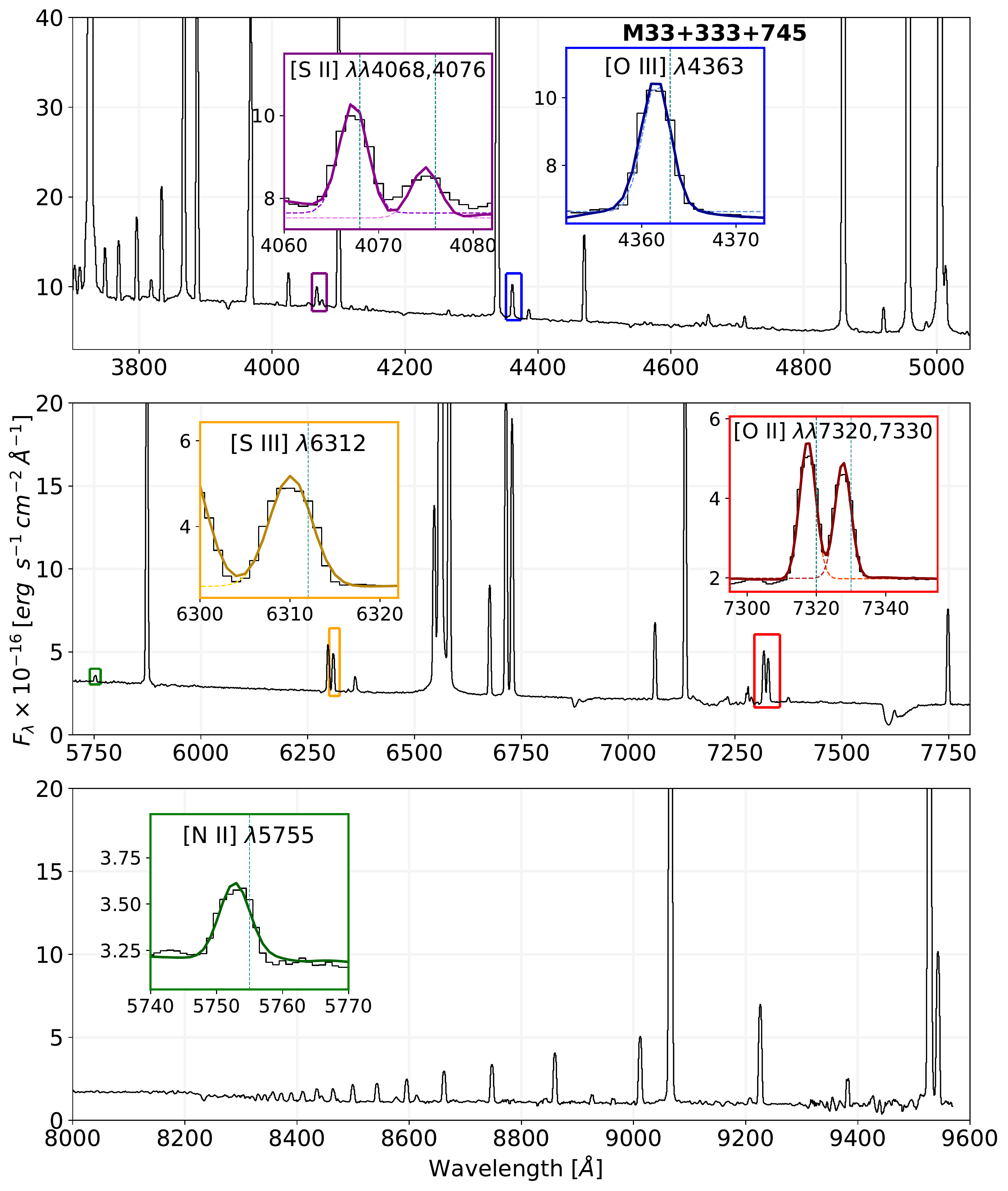}
   \caption{Example MODS spectrum of M33+333+745. Temperature-sensitive auroral lines and their model fits (Gaussian plus STARLIGHT continuum) are provided in the subplots. All five commonly-employed auroral lines are detected at high S/N. While the spectrum has been corrected for the systemic velocity of M33, the rotational velocity of this extended region has shifted the lines blueward of their theoretical wavelengths (vertical dotted blue lines).}
   \label{fig:spectra}
\end{figure*}

Aspects of the MODS data reduction pipeline\footnote{The MODS reduction pipeline was developed by Kevin Croxall with funding from NSF Grant AST-1108693.\ Details at http://www.astronomy.ohio-state.edu/MODS/Software/modsIDL/} are detailed in previous works \citep{berg2015,roge2021}. As such, we only provide a brief summary of the steps taken. The modsCCDRed \textsc{Python} programs \citep{modsCCDRed19} are used to bias subtract and flat field the raw images before combination. Standard star and science images are median combined, then are input into the modsIDL pipeline\footnote{This pipeline operates in the XIDL reduction package, http://www.ucolick.org/$\sim$xavier/IDL/} \citep{modsIDL19}. Standard stars, which are observed on each night of science observation, are processed and used for flux calibration \citep{bohl2014}. Sky subtraction and region extraction are performed on each slit while allowing for multiple sky areas or \hii regions to be extracted within the same slit. Each MOS mask has at least two slits cut away from any prominent ionized gas and which can be used if local sky subtraction is not possible in one of the MOS slits. Finally, the blue and red spectra are re-sampled and combined at 5700 \AA. As an example of the sensitivity and wavelength coverage of MODS, Figure \ref{fig:spectra} plots the high-S/N spectrum of the \hii region M33+333+745. This region has clearly-detected \te-sensitive auroral lines, Balmer and Paschen sequences, and faint emission lines from other metal ions such as \ion{C}{2} and [\ion{Cl}{3}].

The underlying stellar continuum within an \hii region is fit using the STARLIGHTv04\footnote{www.starlight.ufsc.br} spectral synthesis code \citep{starlight} with the stellar population models of \citet{bruz2003}. We mask out strong nebular lines and the area near the dichroic cross-over when fitting the stellar continuum, and we allow for a linear, nebular continuum component in the total fit. The net continuum is subtracted from the spectrum, and the emission lines are fit with Gaussian profiles. The full spectrum is broken into ten wavelength bands in which all Gaussian profiles have the same full width at half maximum (FWHM) and global velocity shift. Line multiplets that are blended at MODS's resolution, such as \oii$\lambda\lambda$3726,3729, are fit using Gaussians with fixed wavelength separation and FWHM.

\vspace{-7mm}
\begin{deluxetable*}{lccccc|lccccc}
\tablecaption{M33 LBT/MODS Observations}
\tablewidth{\textwidth}
\renewcommand{\arraystretch}{0.7}
\tablehead{
   \colhead{\hii}  & 
   \colhead{R.A.}  & 
   \colhead{Dec.}  & 
   \colhead{R$_{g}$}  & 
   \colhead{R$_{g}$/R$_{e}$}  & 
   \colhead{Literature}  & 
   \colhead{\hii}  & 
   \colhead{R.A.}  & 
   \colhead{Dec.}  & 
   \colhead{R$_{g}$}  & 
   \colhead{R$_{g}$/R$_{e}$}  & 
   \colhead{Literature} \vspace{-2ex}\\
   \colhead{Region}  & 
   \colhead{(J2000)}  & 
   \colhead{(J2000)}  & 
   \colhead{(kpc)}  & 
   \colhead{ }  & 
   \colhead{Obs.}  & 
   \colhead{Region}  & 
   \colhead{(J2000)}  & 
   \colhead{(J2000)}  & 
   \colhead{(kpc)}  & 
   \colhead{ }  & 
   \colhead{Obs.}}
\startdata
M33$-$2+25  & 1:33:50.4  & 30:39:55.37  & 0.13  & 0.05  &    & M33$-$224$-$437  & 1:33:33.2  & 30:32:13.23  & 2.07  & 0.89  &   \\
M33+23$-$9  & 1:33:52.4  & 30:39:20.98  & 0.18  & 0.08  &    & M33+253$-$141  & 1:34:10.1  & 30:37:09.18  & 2.09  & 0.90  &   \\
M33$-$23+16  & 1:33:48.8  & 30:39:45.97  & 0.20  & 0.09  &    & M33+19+466  & 1:33:52.1  & 30:47:15.83  & 2.14  & 0.93  &   \\
M33$-$28+6  & 1:33:48.4  & 30:39:35.66  & 0.21  & 0.09  & R08,B11,T16,L17   & M33+263+423  & 1:34:10.9  & 30:46:32.85  & 2.15  & 0.93  &   \\
M33$-$36$-$52  & 1:33:47.8  & 30:38:37.55  & 0.28  & 0.12  & R08,B11,T16   & M33+209+473  & 1:34:06.8  & 30:47:22.42  & 2.16  & 0.93  &   \\
M33+2+111  & 1:33:50.7  & 30:41:20.96  & 0.52  & 0.22  &    & M33+285+399  & 1:34:12.6  & 30:46:08.4  & 2.17  & 0.94  &   \\
M33$-$6$-$120  & 1:33:50.1  & 30:37:30.09  & 0.55  & 0.24  & C06,R08,B11    & M33+221+476  & 1:34:07.7  & 30:47:25.68  & 2.20  & 0.95  &   \\
M33+78+91  & 1:33:56.6  & 30:41:00.69  & 0.55  & 0.24  &    & M33$-$263$-$461  & 1:33:30.2  & 30:31:48.66  & 2.26  & 0.98  & L17  \\
M33$-$89$-$21  & 1:33:43.7  & 30:39:08.91  & 0.59  & 0.25  &    & M33$-$267$-$462  & 1:33:29.9  & 30:31:47.76  & 2.27  & 0.98  &   \\
M33$-$42$-$138  & 1:33:47.3  & 30:37:12.06  & 0.60  & 0.26  &    & M33$-$128+386  & 1:33:40.6  & 30:45:55.93  & 2.29  & 0.99  &   \\
M33+108+25  & 1:33:58.9  & 30:39:55.33  & 0.71  & 0.31  &    & M33$-$143+371  & 1:33:39.5  & 30:45:40.56  & 2.30  & 0.99  & A22  \\
M33$-$66$-$161  & 1:33:45.5  & 30:36:48.59  & 0.73  & 0.31  & R08,B11   & M33+285+456  & 1:34:12.6  & 30:47:05.86  & 2.32  & 1.00  &   \\
M33$-$114$-$123  & 1:33:41.7  & 30:37:26.61  & 0.79  & 0.34  &    & M33+267$-$191  & 1:34:11.2  & 30:36:19.22  & 2.34  & 1.01  &   \\
M33+89+165  & 1:33:57.5  & 30:42:14.43  & 0.79  & 0.34  &    & M33+121$-$405  & 1:33:60.0  & 30:32:44.53  & 2.34  & 1.01  & B11,L17  \\
M33$-$88$-$166  & 1:33:43.8  & 30:36:43.67  & 0.79  & 0.34  &    & M33+300+471  & 1:34:13.8  & 30:47:20.71  & 2.42  & 1.05  &   \\
M33+125+78  & 1:34:00.2  & 30:40:47.65  & 0.81  & 0.35  & R08   & M33+266$-$221  & 1:34:11.2  & 30:35:48.77  & 2.43  & 1.05  &   \\
M33$-$89+89  & 1:33:43.7  & 30:40:58.55  & 0.86  & 0.37  & B11   & M33+299$-$164  & 1:34:13.7  & 30:36:45.89  & 2.46  & 1.07  &   \\
M33$-$144$-$78  & 1:33:39.4  & 30:38:12.2  & 0.93  & 0.40  &    & M33+208+567  & 1:34:06.7  & 30:48:56.99  & 2.52  & 1.09  &   \\
M33$-$150$-$79  & 1:33:39.0  & 30:38:10.55  & 0.97  & 0.42  &    & M33+94+574  & 1:33:57.8  & 30:49:04.34  & 2.53  & 1.09  & C06  \\
M33+70+228  & 1:33:56.0  & 30:43:17.42  & 1.00  & 0.43  &    & M33+107+581  & 1:33:58.9  & 30:49:11.22  & 2.55  & 1.10  & A22  \\
M33+29+261  & 1:33:52.8  & 30:43:51.29  & 1.17  & 0.50  &    & M33+322$-$139  & 1:34:15.5  & 30:37:10.79  & 2.55  & 1.10  & R08,T16,L17,A22  \\
M33+135+264  & 1:34:01.0  & 30:43:53.91  & 1.25  & 0.54  & L17   & M33+119+592  & 1:33:59.8  & 30:49:21.71  & 2.59  & 1.12  &   \\
M33+146+266  & 1:34:01.9  & 30:43:56.26  & 1.29  & 0.56  &    & M33+299+541  & 1:34:13.7  & 30:48:30.9  & 2.62  & 1.13  & A22  \\
M33+143+328  & 1:34:01.7  & 30:44:57.46  & 1.49  & 0.65  &    & M33+334$-$135  & 1:34:16.5  & 30:37:14.7  & 2.62  & 1.13  & L17  \\
M33$-$24$-$333  & 1:33:48.7  & 30:33:56.89  & 1.51  & 0.65  &    & M33+328+543  & 1:34:16.0  & 30:48:33.03  & 2.72  & 1.18  &   \\
M33+113$-$224  & 1:33:59.3  & 30:35:46.33  & 1.52  & 0.66  & B11   & M33+306$-$276  & 1:34:14.2  & 30:34:53.62  & 2.87  & 1.24  &   \\
M33+69+352  & 1:33:55.9  & 30:45:21.87  & 1.54  & 0.67  &    & M33+369+545  & 1:34:19.2  & 30:48:34.82  & 2.88  & 1.25  &   \\
M33+88$-$258  & 1:33:57.4  & 30:35:11.76  & 1.54  & 0.67  &    & M33+380+578  & 1:34:20.0  & 30:49:07.66  & 3.01  & 1.30  &   \\
M33+62+354  & 1:33:55.4  & 30:45:23.73  & 1.55  & 0.67  &    & M33+400+552  & 1:34:21.6  & 30:48:41.86  & 3.02  & 1.31  &   \\
M33$-$36+312  & 1:33:47.8  & 30:44:42.04  & 1.57  & 0.68  &    & M33+405+554  & 1:34:22.0  & 30:48:43.29  & 3.05  & 1.32  &   \\
M33$-$2$-$340  & 1:33:50.4  & 30:33:49.51  & 1.59  & 0.69  &   & M33+298$-$344  & 1:34:13.7  & 30:33:45.39  & 3.06  & 1.32  & R08,L17  \\
M33$-$65+302  & 1:33:45.5  & 30:44:31.99  & 1.64  & 0.71  &    & M33+421+560  & 1:34:23.2  & 30:48:50.12  & 3.13  & 1.35  &   \\
M33$-$108$-$389  & 1:33:42.2  & 30:33:01.36  & 1.70  & 0.73  &    & M33+313$-$342  & 1:34:14.8  & 30:33:48.01  & 3.14  & 1.36  & L17  \\
M33+14$-$355  & 1:33:51.7  & 30:33:34.98  & 1.70  & 0.74  &    & M33+325$-$329  & 1:34:15.7  & 30:34:00.6  & 3.17  & 1.37  &   \\
M33+33+382  & 1:33:53.1  & 30:45:51.7  & 1.72  & 0.74  &    & M33+330$-$345  & 1:34:16.1  & 30:33:44.81  & 3.25  & 1.41  &   \\
M33$-$35$-$385  & 1:33:47.8  & 30:33:04.79  & 1.73  & 0.75  & B11   & M33+345$-$344  & 1:34:17.3  & 30:33:45.44  & 3.34  & 1.45  & R08  \\
M33$-$78+311  & 1:33:44.5  & 30:44:41.0  & 1.73  & 0.75  &    & M33+333+745  & 1:34:16.5  & 30:51:54.32  & 3.41  & 1.47  & C06,L17  \\
M33$-$224$-$346  & 1:33:33.2  & 30:33:43.31  & 1.79  & 0.77  &    & M33+417$-$254  & 1:34:22.9  & 30:35:16.08  & 3.51  & 1.52  &   \\
M33+116$-$286  & 1:33:59.5  & 30:34:43.51  & 1.80  & 0.78  & A22   & M33+371$-$348  & 1:34:19.3  & 30:33:41.27  & 3.52  & 1.52  & R08  \\
M33$-$99+311  & 1:33:42.9  & 30:44:40.53  & 1.83  & 0.79  & A22   & M33+388$-$320  & 1:34:20.6  & 30:34:09.93  & 3.53  & 1.53  &   \\
M33$-$148$-$412  & 1:33:39.1  & 30:32:37.49  & 1.83  & 0.79  &    & M33+541+448  & 1:34:32.6  & 30:46:57.31  & 3.57  & 1.55  & L17  \\
M33$-$196$-$396  & 1:33:35.4  & 30:32:53.73  & 1.86  & 0.80  &    & M33+553+448  & 1:34:33.5  & 30:46:57.07  & 3.64  & 1.57  &   \\
M33+122$-$294  & 1:34:00.0  & 30:34:36.04  & 1.87  & 0.81  &    & M33$-$464+348  & 1:33:14.5  & 30:45:17.71  & 4.12  & 1.78  & L17  \\
M33+26+421  & 1:33:52.6  & 30:46:30.62  & 1.91  & 0.83  &    & M33$-$507+346  & 1:33:11.3  & 30:45:15.59  & 4.38  & 1.90  & C06,T16,A22  \\
M33+46$-$380  & 1:33:54.1  & 30:33:09.67  & 1.92  & 0.83  & R08,B11   & M33$-$72$-$1072  & 1:33:45.0  & 30:21:38.08  & 4.86  & 2.10  & R08,L17,A22  \\
M33+126$-$313  & 1:34:00.3  & 30:34:17.13  & 1.97  & 0.85  & R08,B11,T16   & M33$-$181$-$1156  & 1:33:36.6  & 30:20:13.48  & 5.09  & 2.20  & R08,L17  \\
M33$-$77$-$449  & 1:33:44.6  & 30:32:01.2  & 1.97  & 0.85  & B11,A22   & M33$-$438+800  & 1:33:16.6  & 30:52:49.7  & 5.63  & 2.44  & L17,A22  \\
M33$-$168$-$448  & 1:33:37.6  & 30:32:02.18  & 1.99  & 0.86  &    & M33$-$442+797  & 1:33:16.2  & 30:52:46.13  & 5.64  & 2.44  & A22  \\
M33+175+446  & 1:34:04.1  & 30:46:55.4  & 1.99  & 0.86  &    & M33$-$610$-$1690  & 1:33:03.5  & 30:11:19.06  & 7.49  & 3.24  & C06,L17  \\
M33$-$211$-$438  & 1:33:34.2  & 30:32:12.21  & 2.04  & 0.88  & R08,L17   &    &    &    &    &    &   \\
\enddata
\label{t:m33Obs}
\tablecomments{Compilation of \hii regions CHAOS observed in M33. Columns 1 and 7: \hii region ID given by the R.A. and Decl. offset relative to the center of M33. Columns 2 and 8: R.A. of the \hii region in hours, minutes, and seconds. Columns 3 and 9: Decl. of the \hii region in degrees, arcminutes, and arcseconds. Columns 4 and 10: \hii region distance from the center of M33 in kpc. Columns 5 and 11: \hii region distance from the center of M33 normalized to the effective radius of the galaxy. Columns 6 and 12: Literature studies that have observed the \hii region, see text for shorthand citations (Section 2.3).} 
\end{deluxetable*}

The direct temperature method is exponentially sensitive to the auroral-to-nebular line ratio, requiring utmost care when fitting these lines. The Gaussian fit to each line is checked against a fit of the profile using the IRAF\footnote{IRAF is distributed by the National Optical Astronomy Observatory, which is operated by the Association of Universities for Research in Astronomy, Inc., under cooperative agreement with the National Science Foundation.} \textsc{splot} routine; the flux of the emission line and RMS noise in the continuum are updated to the \textsc{splot} values only when the there is significant disagreement with the modsIDL fitting program. The uncertainty on the emission line flux, adapted from \citet{berg2013} and reported in \citet{roge2021}, is a combination of RMS noise in the continuum around the line profile and a 2\% uncertainty associated with the standard star flux calibration \citep{oke1990}.

The exception to the above fitting routine is for the Balmer lines, which are necessary to measure the line-of-sight reddening for each region. The reddening method that we employ is described in \citet{roge2021} and is a modified version of the technique detailed in \citet{oliv2001} with an MCMC component expanded upon by \citet{aver2021}. In short, we fit a linear function to the continuum across the stellar absorption well, and then fit a Gaussian profile to the Balmer line above the well. Theoretical Balmer line ratios are calculated for \den\ = 10$^2$ cm$^{-3}$ and \te\ = 10$^4$ K from the tables of \citet{stor1995}. We then determine the combination of $c(H\beta)$ and the equivalent width of the stellar Balmer absorption features, $a_H$, that produces agreement between the measured and theoretical H$\alpha$/H$\beta$, H$\gamma$/H$\beta$, and H$\delta$/H$\beta$ ratios. The spectrum is reddening corrected using the best-fit $c(H\beta)$ and the reddening law from \citet{card1989} assuming R$_V$ = 3.1. The electron temperatures are calculated from the resulting line intensities, and new theoretical Balmer line ratios are generated for the determined \te. The process is iterated until convergence (change in \te\ $<$ 20 K).

In Table \ref{t:m33Detect}, we provide the emission line detections for each \hii region. Only emission lines with S/N $>$ 3 are considered detected and, subsequently, used for temperature/abundance analysis. In addition to the auroral line detections, we also indicate the detection of other significant emission lines (Columns 8-11) and the presence of Wolf-Rayet features (Column 12). Table \ref{t:m33Int} in the Appendix provides the emission line intensities, $c(H\beta)$, and $a_H$ for each region. We report the combined flux of \oii$\lambda\lambda$3726,3729 as a single line, \oii$\lambda$3727. We exclude the object M33$-$224$-$346 from the tables in the Appendix and from the following analysis because it is coincident with the Luminous Blue Variable (LBV) M33C-7256 \citep{mass2007,hump2014}. The spectrum of this object is characterized by intense Balmer, \ion{Fe}{2}, and \ion{O}{1} recombination lines while the \oii\ and \oiii\ strong lines are extremely faint. The lack of \oii\ and \oiii\ strong lines would indicate that the bulk of the emission is coming from the LBV and, therefore, the object should not be included in the following \hii region abundance analysis.

\vspace{-9mm}
\begin{deluxetable*}{lcccccc}[ht!]
\tablecaption{M33 Literature Direct Abundance Studies}
\tablewidth{\textwidth}
\tablehead{ 
  \colhead{Reference}	&
  \colhead{$\lambda$ Coverage (\AA)}  &
  \colhead{Resolution}  &
  \colhead{\# H II Regions}	&
  \colhead{\te-Sensitive Lines} &
  \colhead{\te-\te\ } \vspace{-2ex} \\ 
   & & & \colhead{with Direct O/H} & & \colhead{Relations}
  }
\startdata
    \citet{croc2006} (C06)    & 3600--5100   & $\sim$2 \AA   & 6 & \oiii     & [1] \\
    \citet{roso2008} (R08)   & 3600--5400   & $\sim$5 \AA   & 61 & \oiii    & [1] \\
    \citet{bres2011} (B11)   & 3500--5100   & $\sim$5.5 \AA   & 8 & \oiii     & [1] \\
    \citet{tori2016} (T16)    & 3600--7600   & R$\approx$2,500   & 11 & \oiii, \nii, \oii, \sii    & [2] \\
    \citet{lin2017} (L17)     & 3650--9200   & $\sim$6.2 \AA   & 38\tablenotemark{a} & \oiii, \nii  & [1] \\
    \citet{alex2022} (A22)    & 3700--9099   & R$\approx$1,800   & 27    & \oiii, \nii, \oii  & None\\
    This Study                & 3200--10000  & R$\approx$2,000   & 65 & \oiii, \nii, \siii, \oii, \sii    & \S 3.1
\enddata
\label{t:m33lit}
\tablecomments{The compilation of literature abundance studies which obtain \te\ measurements in the \hii\ regions of M33. We focus on studies with many \hii\ regions from which abundance gradients can be measured. The columns provide the following information: reference to the study; the wavelength coverage of the detector used (in \AA); the quoted resolution of the spectra; the reported number of regions with at least one \te-sensitive auroral line detected; the \te-sensitive lines detected in M33; and the \te-\te\ relations applied. The \te-\te\ relations are: [1] \citet{camp1986,garn1992}; [2] N/A unless missing \oiii$\lambda$4363, in which case use empirical relation from \citet{este2009}.
\tablenotetext{a}{More regions have measurements of \oiii\ and \nii\ auroral lines but are rejected based on the shape of the line profile.}}
\end{deluxetable*}
\subsection{Ancillary Data}
Table \ref{t:m33lit} compiles previous literature studies that obtain direct abundances in the \hii regions of M33. While there are many prior abundance studies in M33 (including strong-line and recombination line abundances), we focus on this sample given the wealth of direct abundances and the coverage/overlap of the \hii regions. No two samples in Table \ref{t:m33lit} are directly comparable: each has a different number of regions with \te\ measurements, wavelength coverage, spectral resolution, and auroral lines used.

\subsubsection{Crockett et al. (2006)}

While early direct abundance studies of local galaxies included the bright \hii regions in M33 (see Introduction), \citet[][hereafter C06]{croc2006} is the earliest study with a significant number of direct abundances in M33. \citetalias{croc2006} observed 13 \hii regions with the Mayall Telescope at Kitt Peak National Observatory and obtained optical spectra in the wavelength range 3600--5100 \AA. This wavelength range contains the auroral and strong nebular lines necessary to measure \te\oiii, which was possible in six of the \hii regions. The auroral line \sii$\lambda$4069 was also measured in this range but the strong nebular lines \sii$\lambda\lambda$6717,6731 were not, preventing a measure of \te\sii. Using the nebular lines of \oiii\ and \oii, oxygen abundances were calculated and a gradient was obtained using these six regions and five regions from \citet{vilchez1988}. Additionally, Ne abundances were obtained in these regions using the [\ion{Ne}{3}]$\lambda$3865 line, \te\oiii, and the strong lines of oxygen as an ICF for unobserved ionization states of Ne.

\subsubsection{Rosolowsky \& Simon (2008)}

\citet[][hereafter R08]{roso2008} obtained direct abundances in 61 \hii regions in the southern half of M33 using the Low Resolution Imaging Spectrometer on Keck I, producing one of the largest homogeneous samples of direct abundances in a spiral galaxy. Similar to \citetalias{croc2006}, \te\oiii\ was the only electron temperature measured and subsequently used for direct abundance calculations. \citetalias{roso2008} found that these regions produce a negative O/H gradient but with an intrinsic dispersion about the gradient of $\sigma_{int}$ = 0.11 dex, larger than the uncertainties on the majority of the oxygen abundances. The CHAOS observations target regions across the full disk of M33 and with broader wavelength coverage, allowing for both a comparison of abundances in similar regions and for an assessment of the abundance dispersion that \citetalias{roso2008} measure in the southern half of the galaxy.


\subsubsection{Bresolin (2011)}

To evaluate the magnitude of the intrinsic dispersion in O/H measured by \citetalias{roso2008}, \citet[][hereafter B11]{bres2011} observed 25 central \hii regions in M33 with the Gemini Multi-Object Spectrograph on the Gemini North Telescope. Of the 25 observed (ten of which were also observed by \citetalias{roso2008}), only eight had significant \oiii$\lambda$4363 detections. Using these abundances and direct O/H with low uncertainty from the literature, the scatter in the inner 2 kpc is reduced to $\sim$0.06 dex. The scatter is further explored through strong-line analysis, where regions with high S/N \oiii$\lambda$4363 and acceptable ionization conditions for the \citet{mcGa1991} calibration of $R_{23}$ $=$ log((\oii$\lambda$3727 + \oiii$\lambda\lambda$4959,5007)/H$\beta$) \citep{page1979} produce a strong-line abundance gradient with a dispersion of just 0.05 dex. While the argument for a small intrinsic dispersion about the oxygen abundance gradient is convincing, the direct abundances measured and recalculated from the literature rely on a single \te\ measurement and, therefore, the \te-\te\ relations. An accurate, direct measure of \te\ in all ionization zones is needed to produce robust abundances and to determine if the inferred low-ionization zone \te\ is responsible for some of the direct O/H scatter.

\subsubsection{Toribio San Cipriano et al. (2016)}

\citet[][hereafter T16]{tori2016} observed 11 \hii regions in M33 with the Optical System for Imaging and low-Intermediate-Resolution Integrated Spectroscopy (OSIRIS) instrument on the Gran Telescopio de Canarias to obtain C and O recombination line abundances and the C/O gradient within the galaxy. The broad wavelength coverage allows for \oiii, \oii, and \nii\ temperature calculations in all observed regions, thereby eliminating the need for a \te-\te\ relation to obtain the low-ionization zone temperature. While other literature studies utilize larger samples, \citetalias{tori2016} present the highest spectral resolution data and target central and extended (R$_g$ $\sim$ 7.5 kpc) \hii regions, which enables an understanding of how properties such as \te\nii\ and O/H change over large spatial scales. CHAOS has adopted a similar methodology, allowing for a comparison of these and other properties in many more \hii regions.

\subsubsection{Lin et al. (2017)}

Recently, extensive strong-line abundance studies have been carried out in M33. \citet[][hereafter L17]{lin2017} used the Hectospec fiber system on the Multiple Mirror Telescope to observe 413 \hii regions over a wavelength range of 3650--9100 \AA. Of these, 385 and 38 regions have strong-line and direct abundances, respectively. The oxygen abundances were calculated via \te\oiii\ for the high-ionization zone and the \citet{garn1992} \te-\te\ relation for the low-ionization zone, while \te\nii\ was used for N abundances only. While a comparison between the strong line and direct abundances in many \hii regions is not the focus of this work, this sample includes \hii regions with abundances outside the previously mentioned samples and, therefore, makes for a worthwhile addition.

\subsubsection{Alexeeva \& Zhao (2022)}

\citet[][hereafter A22]{alex2022} report on another strong-line focused abundance study in M33. Their sample includes 110 M33 \hii region spectra from Data Release 7 of the Large sky Area Multi-Object fiber Spectroscopic Telescope (LAMOST) with a wavelength coverage of 3700--9099 \AA\ with a resolution of R$\approx$1800, sufficient to obtain direct temperatures of \oiii, \nii, and \oii. In total, 27 \hii regions have direct abundances including extended objects with CHAOS longslit observations. Both \citetalias{lin2017} and \citetalias{alex2022} measured shallow O/H gradients relative to other literature studies, making an evaluation of these data and the resulting direct temperatures all the more critical to understand the chemical evolution of M33.

\section{Electron Temperatures and ICFs}

\subsection{CHAOS Electron Temperatures}

The MODS spectra obtained in M33 have the wavelength coverage and resolution to measure the electron gas temperature from multiple ions. The temperature-sensitive emission line ratios necessary to measure these temperatures are: \oiii$\lambda$4363/$\lambda\lambda$4959,5007, \nii$\lambda$5755/$\lambda\lambda$6548,6584, \siii$\lambda$6312/$\lambda\lambda$9069,9532, \oii$\lambda\lambda$7320,7330/$\lambda$3727, and \sii$\lambda\lambda$4069,4076/$\lambda\lambda$6717,6731. The temperature measured from each ion corresponds to the gas within the \hii region which contains that ion. For example, O$^{2+}$ is present in an \hii region only where the average degree of ionization is high, and so \te\oiii\ corresponds to the temperature from this high-ionization zone. We use a three-zone model for each \hii region: the ions N$^+$, O$^+$, and S$^+$ are present in the low-ionization zone; the intermediate-ionization zone contains S$^{2+}$, Cl$^{2+}$, and Ar$^{2+}$; the gas responsible for the emission of O$^{2+}$, Ne$^{2+}$, and Ar$^{3+}$ is the high-ionization zone. The strength of having a large wavelength coverage is the capability of measuring direct electron temperatures from all of these ionization zones, allowing for reliable ionic abundance measurements. 

The electron density can be calculated from the \sii$\lambda$6717/\sii$\lambda$6731 and [\ion{Cl}{3}]$\lambda$5517/[\ion{Cl}{3}]$\lambda$5537 intensity ratios. Provided the lower ionization energy required to produce S$^+$ and the scarcity of Cl in the ISM, it is significantly easier to detect and use the \sii\ strong lines as density diagnostics. Despite the lower intensity and abundance of Cl, we detect the [\ion{Cl}{3}]$\lambda\lambda$5517,5537 pair in 20 \hii regions (see Table \ref{t:m33Detect}).

To calculate \te\ from the significantly-detected auroral lines, we use the above line intensity ratios as input into the \textsc{Python} \textsc{PyNeb} \citep{luri2012,luri2015L} package's \textsc{getTemDen} function. The temperature-sensitive line ratios have a weak dependence on the electron density, but in the low-density limit (n$_e <$ 10$^2$ cm$^{-3}$) the ratios are essentially density-independent. Densities are calculated using the \sii$\lambda$6717/$\lambda$6731 and [\ion{Cl}{3}]$\lambda$5517/$\lambda$5537 intensity ratios and low- and intermediate-ionization zone temperatures, respectively, in the same \textsc{getTemDen} function. The majority of the \hii regions in M33 have a measured n$_e$(\sii) consistent with the low-density limit. As such, temperatures are calculated using the \textsc{getTemDen} function with the emission line intensity ratios and at a fixed electron density of 10$^2$ cm$^{-3}$.

We use a MCMC approach to obtain the uncertainties on the electron temperatures. We generate a distribution of 1000 auroral-to-nebular line ratios assuming a normal distribution centered on the measured line ratio and with standard deviation equal to the uncertainty on the ratio. We then calculate the temperature from all ratios in the distribution and take the standard deviation of these values as the uncertainty on the measured temperature. During this process, we fix the electron density to n$_e$ = 10$^2$ cm$^{-3}$. The same MCMC technique is used for the uncertainty on n$_e$(\sii) and n$_e$([\ion{Cl}{3}]) using the uncertainty on the measured line ratios of \sii\ and [\ion{Cl}{3}] and fixing \te\ to the low- and intermediate-ionization zone values, respectively.

There are a few cases where additional care is taken with the line ratios used for temperature analysis. First, the strong lines \siii$\lambda\lambda$9069,9532 can be contaminated by telluric absorption; a contaminated strong line produces a larger auroral-to-nebular line ratio and, therefore, \te\siii\ than is actually present in the gas. To assess contamination, we compare the measured \siii\ strong line ratios to the theoretical value of \siii$\lambda$9532/$\lambda$9069 = 2.47. The theoretical line ratio is determined from the ratio of \siii$\lambda$9532 and \siii$\lambda$9069 emissivities using the atomic transition probabilities of \citet{froe2006}; it has been shown that different atomic data produce slight variations in this theoretical ratio, but the \citet{froe2006} transition probabilities are consistent with the range of values predicted by the majority of the available datasets \citep[from 2.47 to 2.54, see][]{mend2022b}.

If the measured line ratio agrees with the theoretical ratio within uncertainty, then we use both lines in \te\siii\ calculation. If the measured ratio is greater than theoretical, as is the case for a more contaminated \siii$\lambda$9069, we only use the \siii$\lambda$9532 line, and vice versa for a measured ratio that is less than theoretical. Telluric absorption bands can affect the transmission of large portions of the NIR \citep[see Figure 3 in][]{noll2012}, which could result in contamination of both \siii\ strong lines. While the systemic velocity of M33 is not sufficient enough to blueshift these lines completely away from the absorption bands, the low dispersion in \te\siii\ observed in other CHAOS galaxies \citep{crox2016,berg2020,roge2021} would indicate that this approach is not introducing unphysical scatter in the sulfur temperatures. This technique is repeated for the \oiii$\lambda\lambda$4959,5007 strong lines, although the \oiii$\lambda$5007/$\lambda$4959 ratio agrees with theoretical \citep[2.89 from the ratio of emissivities and the atomic transition probabilities of][]{froe2004} for all regions with significant \oiii$\lambda$4363 auroral line detections.

Second, it is not uncommon for H$\alpha$ to have wing profiles that extend beyond the FWHM of the Gaussian fit. Depending on the region, these wings can blend with \nii$\lambda$6548, which is only 15 \AA\ away from the H$\alpha$ line center at 6563 \AA. The other strong line, \nii$\lambda$6584, is farther from H$\alpha$ and is not as blended with the wing profiles in the majority of the regions. As such, we only use the \nii$\lambda$6584 line to calculate \te\nii\ in the \hii regions of M33.

Lastly, previous studies have reported on possible contamination of \oiii$\lambda$4363 due to the presence of [\ion{Fe}{2}]$\lambda$4360 \citep{curt2017}. For low-resolution spectrographs, the inability to distinguish the \oiii\ and [\ion{Fe}{2}] line may lead to an erroneously high flux measurement for \oiii$\lambda$4363 which produces a temperature that is too large for the high-ionization zone. The presence of [\ion{Fe}{2}]$\lambda$4360 that is comparable to \oiii$\lambda$4363 might be expected for higher-metallicity sources \citep[see discussion in][]{curt2017}, but fluorescence can populate some of the energy levels, including the $^6S_{5/2}$ level that produces [\ion{Fe}{2}]$\lambda$4360, and cause enhanced [\ion{Fe}{2}] emission \citep{rodr1999}. The two lines are slightly blended at the resolution of MODS; while we have previously used the FWHM of \oiii$\lambda$4363 and other neighboring lines to simultaneously fit a second Gaussian at 4360 \AA\ \citep{berg2020,roge2021}, we can also use other [\ion{Fe}{2}] lines originating from the same $^6S_{5/2}$ level to assess the contamination of \oiii$\lambda$4363 by [\ion{Fe}{2}]. From the atomic data of \citet{baut2015}, the emissivity ratio of [\ion{Fe}{2}]$\lambda$4288, which also originates from the $^6S_{5/2}$ level, to [\ion{Fe}{2}]$\lambda$4360 is 1.37. As such, if we do not observe [\ion{Fe}{2}]$\lambda$4288 then the contamination of the \oiii\ auroral line by [\ion{Fe}{2}]$\lambda$4360 is negligible and is not corrected. If the profile of [\ion{Fe}{2}]$\lambda$4288 can be fit (as is the case for four regions), then the corresponding inferred intensity of [\ion{Fe}{2}]$\lambda$4360 is subtracted from \oiii$\lambda$4363 before temperature calculations.

This approach relies on the significant detection of [\ion{Fe}{2}]$\lambda$4288, but even a weak, inferred [\ion{Fe}{2}]$\lambda$4360 could bias faint \oiii$\lambda$4363 such that it becomes a significant detection and produces an unreasonably high \te\oiii. When verifying the fits to the auroral lines, we also check the line profile of \oiii$\lambda$4363 and, if possible, attempt to fit a second Gaussian profile at 4360 \AA\ when the profile is asymmetric. We complete this fit with another strong line in the spectrum to constrain the FWHM of the Gaussians at 4360 \AA\ and 4363 \AA. If [\ion{Fe}{2}]$\lambda$4288 is undetected and if the intensity of \oiii$\lambda$4363 significantly changes when considering a contaminating line at 4360 \AA, we adopt the fit to \oiii$\lambda$4363 from the deblended Gaussian. With these steps, we have attempted to reasonably account for [\ion{Fe}{2}] contamination in a way that is physically (using the intensity of [\ion{Fe}{2}]$\lambda$4288) and observationally (using the symmetry of the Gaussian profile) motivated. 

\begin{figure*}[pt]
\epsscale{1.0}
   \centering
   \includegraphics[width=0.77\textwidth, trim=30 0 30 0,  clip=yes]{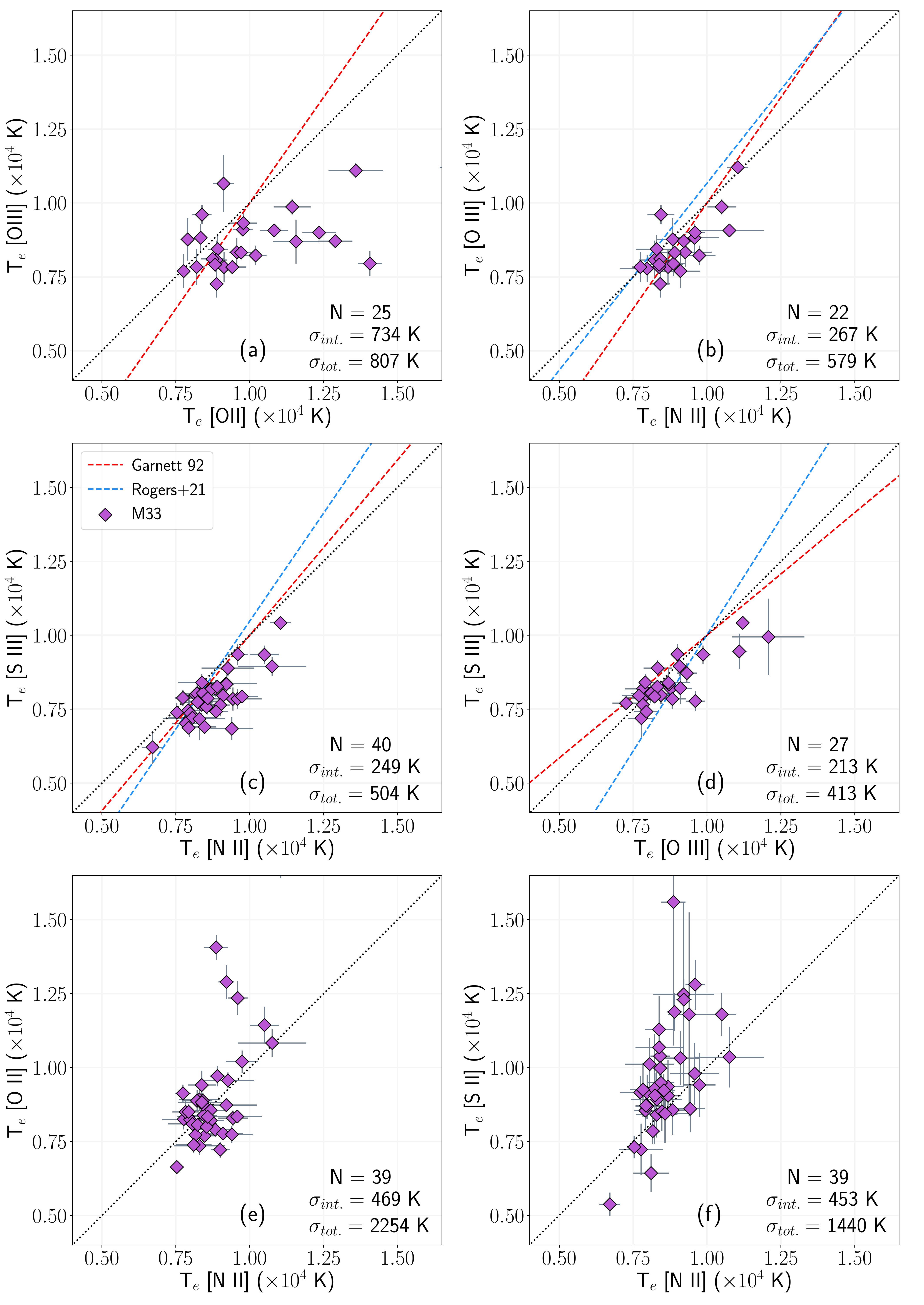}
   \caption{The CHAOS-measured \te-\te\ trends in the \hii regions of M33. Each point represents an \hii region containing a direct \te\ from the corresponding ion. The dashed red and blue lines are the \te-\te\ relations of \citet{garn1992} and \citet{roge2021}, respectively. The number of regions with both direct temperatures, the intrinsic and total scatter in the dependent \te\ are provided in the lower right. \textit{(a) and (b):} The high- vs. low-ionization zone temperatures, the former as measured by \te\oiii\ and the latter measured by \te\oii\ (a) and \te\nii\ (b). \textit{(c) and (d):} Intermediate-ionization zone temperatures from \te\siii\ as compared to \te\nii\ (c) and \te\oiii\ (d). \textit{(e) and (f):} A comparison of the low-ionization zone temperatures measured by \te\nii\ to the temperatures in the same zone as measured from \te\oii\ (e) and \te\sii\ (f).}
   \label{fig:ttrelations}
\end{figure*} 

With the above considerations, we determine the electron temperatures in M33 and report them in Table \ref{t:m33Abun} in the Appendix. Given the dependence of the electron temperature on the abundance of heavy elements, it is expected that temperatures from different ionization zones are related. When a direct temperature is not measured in an ionization zone, these \te-\te\ relations provide a method to infer the ionization zone temperature from the available temperature data. Many previous studies have derived the functional form of \te-\te\ relations from empirical direct temperature data \citep[e.g.,][]{este2009,crox2016,arel2020,roge2021} and from photoionization models \citep[e.g.,][]{camp1986,garn1992,izot2006}. Many of the \hii regions in M33 have multiple direct temperatures, but the size of the sample is not sufficient to derive statistically-significant empirical \te-\te\ relations. Instead, we can compare the direct temperature trends in this galaxy to existing relations.

Figure \ref{fig:ttrelations} plots the M33 direct temperature trends from numerous ions and \hii regions. Each panel plots the direct temperature from one ion against the temperature from another ion measured in the same region, and each temperature is normalized by 10$^4$ K. Reported in the lower right of each panel is the number of \hii regions with both direct temperatures, the intrinsic dispersion, defined as the random scatter in the dependent variable about a linear regression fit to the data, and the total dispersion in \te\ \citep[both of which are calculated using the same technique as][]{bedr2006}. While the intrinsic dispersion depends on which temperature is assumed to be the dependent variable in the linear relation \citep[see discussion in][]{roge2021}, we only provide one permutation of the relations because we are not attempting to derive robust relations from the M33 regions alone. We will report on the global direct temperature trends of the CHAOS sample in a future work.

Plotted as a red dashed line in panels (a) and (b) of Figure \ref{fig:ttrelations} is the commonly applied low-to-high ionization zone \te-\te\ relation of \citet{camp1986,garn1992}. In the majority of M33 abundance studies with only \te\oiii, this is the relation applied to obtain the low-ionization zone temperature (see Table \ref{t:m33lit}). As other studies report \citep{kenn2003b,este2009,berg2015,yate2020}, the trend between the direct temperatures of \te\oiii\ and \te\oii\ is not clear and the temperatures are often scattered over many thousands of K. This is true in our sample of temperatures, where panel (a) shows the intrinsic dispersion is significantly larger than the other temperature trends. \te\nii, another representative temperature of the low-ionization zone, does not show this scatter relative to \te\oiii. In fact, the direct temperatures measured here follow the \citet{camp1986,garn1992} relation well and exhibit small intrinsic scatter.

The observed \te\oiii\ in M33, and in many galaxies, have an effective lower limit that is set by the average electron energy within the nebula: the electron energy needed to excite O$^{2+}$ to the 4th excited state is relatively large, requiring either high \te\ or a relatively bright region to detect \oiii$\lambda$4363. This prevents detections of low \oiii$\lambda$4363 in all but the brightest regions, which is why there are fewer regions with concurrent \te\oiii\ and \te\nii\ in the low-\te\ regime ($\lesssim$ 7500 K). The electron energies required to enable the transitions that produce \nii$\lambda$5755 and \siii$\lambda$6312 are significantly lower, resulting in the larger number of regions with simultaneous \te\nii-\te\siii\ and with \te\ $<$ 7500 K in panel (c).

\citet{garn1992} also report a photoionization model \te-\te\ relation for the intermediate-ionization zone; the extrapolation of the high-to-intermediate relation to the low-ionization zone is plotted in panel (c). Again, there is generally good agreement between the measured \te\nii-\te\siii\ trends in M33 and this relation. The measured temperatures have a small intrinsic dispersion, but the best-fit relation deviates from the relation derived from the other CHAOS galaxies (blue dashed line) at high \te. The relation reported in \citet{roge2021} makes use of 108 direct \te\nii\ and \te\siii, many of which are at lower average \te\ than the \te\ measured in M33 \citep[particularly the temperatures from the high-metallicity galaxies NGC 5194 and NGC 3184, see][]{crox2015,berg2020}. It is not surprising, then, that the temperatures measured here deviate from the trends at the extreme temperature ranges when we do not measure such temperatures in M33.

Panel (d) plots the \te\siii-\te\oiii\ temperatures observed in 27 regions with the photoionization \citep{garn1992} and CHAOS empirical \te\siii-\te\oiii\ relations. The intrinsic dispersion about the M33 relation is a factor of five smaller than that reported in \citet{roge2021}. Again, the sample of regions examines a relatively small area in \te-\te\ space when compared to the full CHAOS \hii region sample, but the large \siii\ temperatures measured in other spiral galaxies \citep[like M101,][]{crox2016} are simply not observed in M33. For many regions at moderate \te, the CHAOS relations describe the \te\ trends relatively well.

Finally, panels (e) and (f) focus on the low-ionization zone temperatures from \nii, \oii, and \sii. Provided that these ions have roughly the same ionization energies, it is expected that their temperatures should be, generally, in agreement. This is roughly the case for the 39 regions with concurrent \te\nii-\te\oii\ and \te\nii-\te\sii\ where most regions are scattered about the 1-to-1 line. This scatter has been measured in other galaxies \citep{este2009,roge2021}, and it might be related to factors such as dielectronic recombination \citep{rubi1986,liu2001}, contamination in the NIR, or the presence of higher temperatures in the gas surrounding the \hii region (the photodissociation region, or PDR). As such, we deprioritize \te\oii\ and \te\sii\ in the following abundance analysis due to the scatter relative to \te\nii.
Note that there are actually more \te\oii\ measurements than any other temperature
(see Table \ref{t:m33Detect}).  A better understanding of the behavior of 
\te\oii\ would represent a significant increase in diagnostic power.

Following \citet{roge2021}, we use the weighted-average electron temperature prioritization when calculating the temperature to use in the low-, intermediate-, and high-ionization zones. This method makes use of the available \nii, \siii, and \oiii\ temperatures in a single region for a robust estimate of the temperature in each ionization zone. With these three temperatures, one can measure a dominant temperature and two inferred temperatures from available \te-\te\ relations, then combine these in a weighted average. This method prioritizes significant \te\ from the dominant ion (i.e., those that have the lowest errors in that ionization zone), or a combination of the inferred temperatures when the dominant ion is not detected or is measured at low S/N. We apply the \te-\te\ relations from \citet{roge2021}, which are calibrated from the CHAOS sample of direct temperatures and employ the intrinsic scatter about the relations to better account for the uncertainties on the inferred temperatures \citep[see discussion in][]{roge2021}. If a region does not have significant \nii, \siii, or \oiii\ auroral line detections, then that region is rejected from the abundance analysis. T$_{e, Low}$, T$_{e, Int}$, and T$_{e, High}$ determined from the weight-average prioritization are reported in Table \ref{t:m33Abun}. The following abundance analysis has been repeated using the standard ionization-based temperature prioritization (i.e., use a single direct temperature if it is the dominant ion in the ionization-zone), and all results are consistent within uncertainty.

\subsection{ICFs for Various Ionic Species}

\vspace{-4mm}
\begin{deluxetable}{lccc}  
\tablecaption{CHAOS Atomic Data}
\tablewidth{\columnwidth}
\tablehead{ 
  \colhead{Ion}	&
  \colhead{Transition Probabilities}	&
  \colhead{Collision Strengths}
  }
\startdata
N$^+$ 	& \citet{froe2004}  &  \citet{taya2011}  \\ 
O$^+$ 	& \citet{froe2004}  &  \citet{kisi2009} \\
O$^{2+}$ 	& \citet{froe2004}  &   \citet{stor2014} \\
Ne$^{2+}$   & \citet{froe2004}  &   \citet{mcLa2000}   \\
S$^+$   & \citet{irim2005}  & \citet{taya2010} \\
S$^{2+}$    & \citet{froe2006}  &   \citet{huds2012}   \\
Cl$^{2+}$   & \citet{rynk2019}   & \citet{butl1989}   \\
Ar$^{2+}$   &   \citet{mend1983}    &   \citet{muno2009}  
\enddata
\label{t:atomic}
\tablecomments{Atomic data used for the calculation of ionic abundances in M33.}
\end{deluxetable}

With the temperatures in each ionization zone and the intensities of the strong lines available in the optical, we can directly calculate the abundance of different ionic species within each zone via the emissivity of the transition that produces the strong line. The \textsc{PyNeb getIonAbundance} function performs this calculation, where a five-level atom model \citep{dero1987} is assumed and the atomic data (transition probabilities and collision strengths) are provided in Table \ref{t:atomic}. To calculate the ionic abundance uncertainty, we propagate the uncertainty on the ionization-zone temperature through the emissivity ratio, then use the resulting emissivity and intensity ratio uncertainty to calculate the total percent error on the ionic abundance. To calculate the total abundance of a given element, it is necessary to obtain an abundance measurement of all ionic species of that element within the \hii region. This may not be possible if a specific ionization state has no observable emission lines in the optical, so ICFs are used to account for the unobserved ionic species.

The total abundance of oxygen within an \hii region is well-described by O = O$^+$ + O$^{2+}$, where it is assumed that there is little neutral O in the \hii region due to the coincident ionization energies of H and O. In very highly ionized regions, oxygen can triply ionize; while there are no observable lines of O$^{3+}$ in the optical, significant \ion{He}{2}$\lambda$4686 indicates the presence of a very-high ionization zone that contains He$^{2+}$, O$^{3+}$, and other ions \citep{berg2021}. In the regions of M33, only one region, M33$-$211$-$438, has a significant detection of narrow \ion{He}{2}$\lambda$4686; all other regions with significant fits to \ion{He}{2}$\lambda$4686 are dominated by the stellar continuum (no clear line, but absorption at 4686 \AA\ predicted by STARLIGHT) or have broad \ion{He}{2}$\lambda$4686 indicative of WR stars. While we could account for the missing O$^{3+}$ in M33$-$211$-$438, we adopt the common assumption that all gas-phase O is in the O$^+$ or O$^{2+}$ states to remain consistent with the majority of the sample. N$^+$ is the only observable ionization state of nitrogen in the optical. N$^+$ and O$^+$ characterize the low-ionization gas in an \hii region, and the ratio of N/O is not expected to change between different ionization zones. The common ICF for N is, therefore, ICF(N) = O/O$^+$ such that N/H = N$^+$/H$^+ \times$ ICF(N). The use of this ICF produces N/O abundances that are found to be within 10\% of the true N/O values from photoionization models \citep[see][]{nava2006,amay2021}.

The ICFs applied for the $\alpha$-element abundances in the CHAOS sample were discussed in \citet{roge2021}, but these ICFs come from an array of sources \citep[for example,][]{peim1969,thua1995,izot2006}. Recently, \citet{amay2021} published new ICFs for Ne, S, Cl, and Ar generated from the Mexican Million Models database \citep[3MdB,][]{mori2015}. The models used for calibrating the ICFs are selected for their ability to reproduce the line ratios observed in blue compact galaxies and giant \hii regions, making them suitable for the sample of regions in M33. Each ICF is a fifth-order polynomial in O$^{2+}$/O and has an associated uncertainty that is itself a polynomial function of O$^{2+}$/O. In prior CHAOS studies it was assumed that the applied ICF had an uncertainty of 10\%, but the uncertainty in a given ICF can vary significantly as a function of O$^{2+}$/O \citep[see Figure 7 in][]{amay2021}. To accurately calculate the uncertainty in the ICFs, make use of the most recent ICFs, and have a consistent source for all $\alpha$-element ICFs, we use the Ne, S, Cl, and Ar ICFs and their uncertainties from \citet{amay2021}. These ICFs are calibrated to calculate $\alpha$/O abundances, and so we use the ionic oxygen abundances, when necessary, to convert to $\alpha$/H abundances.

For the first time in a CHAOS abundance study, we determine Cl$^{2+}$ abundances from a significant number of regions (20 total) using the [\ion{Cl}{3}]$\lambda\lambda$5517,5537 lines. To determine the ionic abundances, we use the calculated n$_e$([\ion{Cl}{3}]), the [\ion{Cl}{3}]$\lambda$5517 intensity, and the intermediate-ionization zone temperature. We require a measure of n$_e$([\ion{Cl}{3}]) to calculate the Cl$^{2+}$ abundance because the assumption of n$_e$ = 10$^2$ cm$^{-3}$ produces different Cl$^{2+}$/H$^+$ when using either [\ion{Cl}{3}]$\lambda$5517 or [\ion{Cl}{3}]$\lambda$5537 individually. However, the energy levels that produce the [\ion{Cl}{3}] density-sensitive lines have higher critical densities than the commonly used \sii\ lines; as such, the emissivity ratio of the two lines is a slowly varying function of n$_e$ in the low density regime that describes the regions of M33 (as evidenced by the \sii\ lines). In fact, four regions have measured line ratios that are not predicted at the provided \te\ in that their ratio is at or above the ratio predicted at n$_e \approx$ 0 cm$^{-3}$. For this same reason, the uncertainties calculated using the MCMC method can be particularly large. Taken together, when the line ratio does not permit a calculation of n$_e$([\ion{Cl}{3}]), we calculate Cl$^{2+}$/H$^+$ from both lines assuming n$_e$ = 10$^2$ cm$^{-3}$ and average the resulting values. We have verified that using I([\ion{Cl}{3}]$\lambda$5517)$+$I([\ion{Cl}{3}]$\lambda$5537) and assuming a density of n$_e$ = 10$^2$ cm$^{-3}$ produces ionic abundances that are consistent with the ionic abundances determined from the above method. The Cl$^{2+}$/O$^{2+}$ abundance is then used with the \citet{amay2021} ICF to determine the Cl/O relative abundance.

\section{Abundance Gradients and Dispersions}

Using the ionic abundances and ICFs described in the previous section, we calculate the total abundances of numerous elements in M33. We first examine the homogeneous abundances determined from the CHAOS observations to assess the magnitude of the intrinsic dispersion in abundances in M33. In the next section we compare these abundances to the measured literature values and merge the samples that we believe to be of similar quality.

\subsection{Oxygen}

The third most abundant element, oxygen is a tracer of high-mass star formation. In many star-forming spiral galaxies, the oxygen abundance is observed to be largest near the center of the galaxy and lowest at the edges of the spiral arms. The negative oxygen abundance gradient in M33 measured by \citetalias{croc2006}, \citetalias{roso2008}, \citetalias{bres2011}, and \citetalias{tori2016} all support the inside-out chemical evolution of this galaxy. The dispersion about the negative abundance gradient is related to processes within the galaxy such as radial motion along spiral arms, local contamination due to pristine gas infall, etc. The scatter measured in M33 from a sample of direct abundances alone ranges between 0.06 dex as measured in the inner 2.2 kpc by \citetalias{bres2011} to 0.11 dex reported by \citetalias{roso2008}.

\begin{figure}[t]
\epsscale{1.0}
   \centering
   \includegraphics[width=0.45\textwidth, trim=40 0 40 0,  clip=yes]{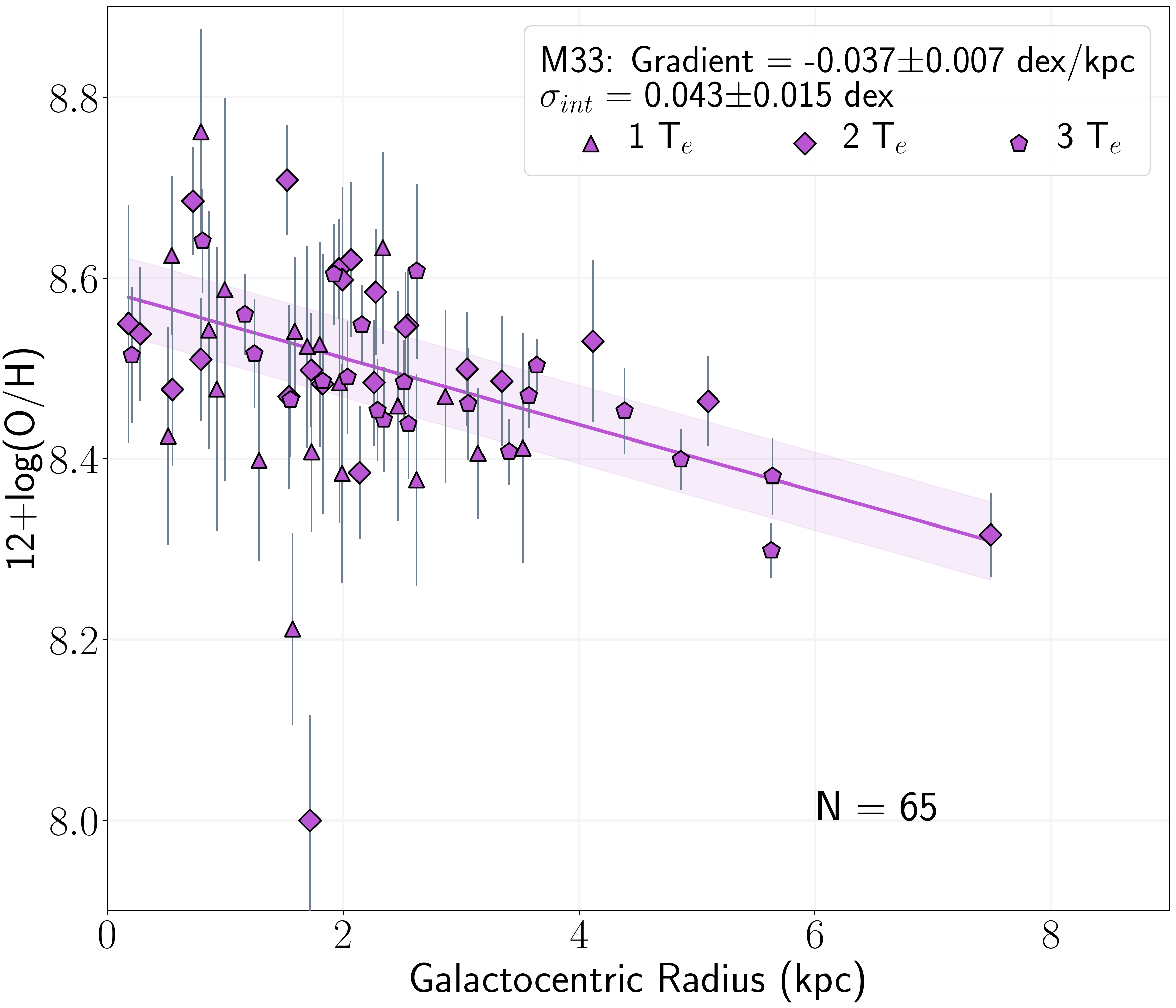}
   \caption{The direct oxygen abundance gradient in M33 measured by CHAOS. The oxygen abundance, 12+log(O/H) (dex), of each region is plotted against the galactocentric radius (in kpc) of the region. The shape of the point represents the number of direct temperatures used to calculate the abundances within the region, as indicated in the legend. The best-fit gradient and intrinsic scatter in O/H about the gradient (provided in the legend) are represented as a solid line and shaded area about the line, respectively.}
   \label{fig:oh_gradient}
\end{figure} 

In Figure \ref{fig:oh_gradient}, we plot the oxygen abundances (in dex) measured in this study against the galactocentric radius of each region in kpc. The linear gradient is fit using the \textsc{Python LINMIX} package\footnote{ https://github.com/jmeyers314/linmix}, which fits a linear function between two variables while considering the uncertainty in each variable and returns the random scatter in the dependent variable about the linear fit \citep[this is the implementation of the fitting program outlined in][]{kell2007}. The shape of each point in Figure \ref{fig:oh_gradient} represents the number of direct temperatures used in the weighted-average ionization-zone temperature calculations: triangles indicate regions with a single direct temperature, diamonds represent regions with two, and pentagons are regions with temperatures from \nii, \siii, and \oiii. The shaded region about the gradient represents the intrinsic dispersion (in dex) about the abundance gradient. The gradient and intrinsic dispersion are provided in the legend, and the number of \hii regions used in fitting the gradient is found in the lower right corner. The O/H gradient in M33 reported here is measured from 65 \hii regions, making this the largest, homogeneous sample of direct abundances in M33.

The gradient measured from this sample is:
\begin{equation}
\mbox{12+log(O/H)}=8.59(\pm0.02) - 0.037(\pm0.007)\times R_{g,kpc}
\label{eq:oh_grad}
\end{equation}
where R$_{g,kpc}$ is the galactocentric radius in kpc. The intrinsic dispersion in oxygen abundance about this gradient is $\sigma_{int} = 0.043\pm0.015$ dex. When normalizing the positions to the effective radius of M33, the gradient is $-$0.081$\pm$0.017 dex/R$_e$. From Eq. \ref{eq:oh_grad}, we confirm the existence of a negative oxygen abundance gradient in M33. This abundance gradient appears to accurately describe the O/H trends from within $\sim$0.13 kpc to the outer \hii regions at nearly 7.5 kpc.

There is non-negligible scatter in the oxygen abundances, with some regions scattered to relatively low ($\lesssim$ 8.2) and high ($\gtrsim$ 8.7) O/H. However, the scattered regions are primarily those with a single electron temperature used to infer the temperature in all ionization zones; while the uncertainty on O/H is reflective of the missing temperatures, the scatter these regions exhibit highlights the importance of having: 1. High S/N measurements of the \te-sensitive auroral lines, and 2. Multiple direct temperatures spanning the typical ionization zones of a region.

The measured intrinsic dispersion is less than the value of 0.11 dex reported by \citetalias{roso2008} and in fairly good agreement with the scatter of 0.06 dex obtained in the inner 2.2 kpc as measured by \citetalias{bres2011}. In fact, the average error in O/H of regions with all three commonly-used electron temperatures is $\langle\delta$(O/H)$_{3}\rangle =$ 0.053 dex, which matches $\sigma_{int}$ within uncertainty. The average O/H uncertainty for regions with two and one available temperature are $\langle\delta$(O/H)$_{2}\rangle =$ 0.078 dex and $\langle\delta$(O/H)$_{1}\rangle =$ 0.118 dex, respectively. Given that intrinsic dispersion in the oxygen abundances is comparable to the average uncertainties of our best measured targets, we find no evidence of significant azimuthal abundance variations in M33.

We examine the distribution of O/H about the best-fit gradient to further explore this claim. If we take $\eta$ to be the difference between the measured and predicted oxygen abundance, where the latter is calculated from Eq. \ref{eq:oh_grad} and the radius of the \hii region, we can determine its average, $\langle\eta\rangle$, and standard deviation, $\sigma(\eta)$, for the regions with one, two, and three direct temperatures used in the weighted average approach. $\langle\eta\rangle$ examines how the abundances in regions that are missing direct temperatures compare to the best-fit gradient, while $\sigma(\eta)$ provides insight on the scatter in O/H in each sample of regions. We find: for regions with three direct temperatures, $\langle\eta_3\rangle =$ 0.00 dex and $\sigma(\eta_3) =$ 0.05 dex; those with two direct temperatures, $\langle\eta_2\rangle =$ 0.03 dex and $\sigma(\eta_2) =$ 0.10 dex; and for regions with a single temperature, $\langle\eta_1\rangle = -$0.04 dex and $\sigma(\eta_1) =$ 0.11 dex. It is not surprising that the average offset from the gradient is very small for the sample with three direct temperatures, as the uncertainty in O/H is relatively small and the best-fit gradient gives these regions a higher priority in the fit. $\sigma(\eta_3)$ is consistent with $\langle\delta$(O/H)$_{3}\rangle$ and $\sigma_{int}$, supporting the claim that the O/H observed in the regions with the highest S/N \te\ spanning multiple ionization zones do not show large azimuthal variations.

Our data show evidence that when a single temperature is measured, requiring the use of \te-\te\ relationships to infer the temperature in the other ionization zone, then the dispersion in the total oxygen abundance is inflated because of the inadequacy of simple \te-\te\ relationships to accurately infer temperatures. The temperature most often inferred is the high-ionization zone temperature, as \oiii$\lambda$4363 is detected in 28 regions (see Table \ref{t:m33Detect}). The weighted average approach utilizes the measured \te\nii\ and/or \te\siii\ with the chosen \te-\te\ relations to infer a high-ionization zone temperature, but the \te\oiii-\te\siii\ relation of \citet{roge2021} shows very large scatter and the \te\oiii-\te\nii\ relation is constructed with a relatively small number of \hii regions. A direct measure of \te\ within this ionization zone is crucial for reliable measurements of the O$^{2+}$ abundance.

Regions with a single electron temperature rely most on the accuracy of the \te-\te\ relations. \citet{pere2003} highlight this weakness in direct abundance measurements, and \citet{arel2020} find that applying a single \te\oiii\ or \te\nii\ with various \te-\te\ relations can produce differences in total N and O abundances greater than 0.2 dex, especially so when \te\oiii\ is inferred from \te\nii. While the larger errors on their direct O/H reflect the uncertainty in the relations, the true temperature in a given region/ionization zone can deviate from the simple, linear relations, resulting in erroneous temperatures/abundance trends. Therefore, a measure of the true intrinsic dispersion in O/H within a spiral galaxy can only be obtained with direct measures of the low- and high-ionization zone temperature; the use of inferred temperatures can produce abundances that generally agree with the abundance gradient but show enhanced scatter that is not reflected in the most reliable measurements of \te\ and O/H.

\subsection{Nitrogen}

\begin{figure}[t]
\epsscale{1.0}
   \centering
   \includegraphics[width=0.45\textwidth, trim=50 0 40 0,  clip=yes]{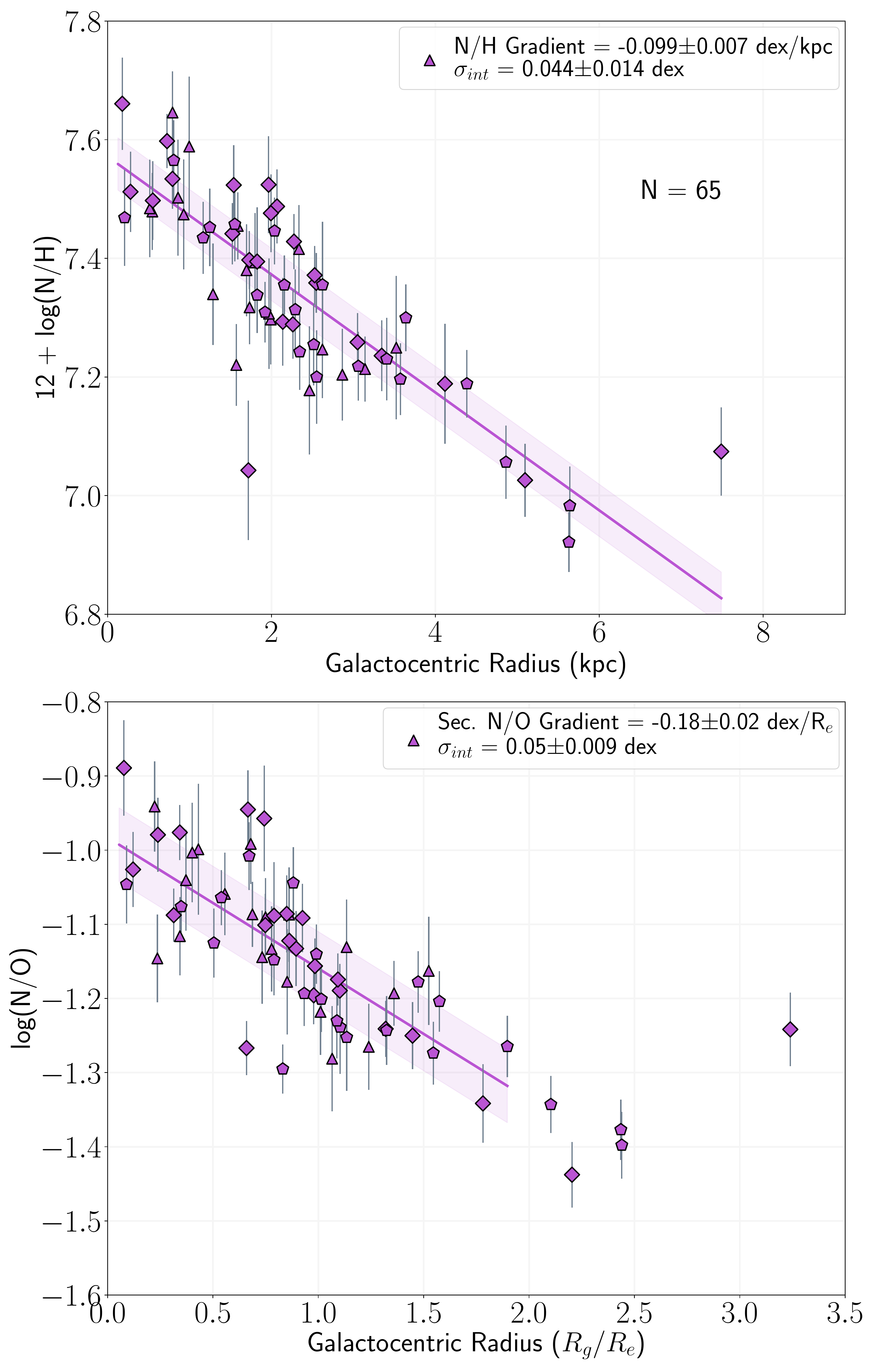}
   \caption{The direct N abundance gradients in M33. In each panel, the gradient and intrinsic scatter are provided as the solid line and shaded regions, respectively, and are reported in the plot legends. The symbols follow the same representation as Figure \ref{fig:oh_gradient}. \textit{Top Panel:} 12+log(N/H) as a function of R$_{kpc}$. \textit{Bottom Panel:} log(N/O) relative abundances as a function of R$_g$/R$_{e}$, or the galactocentric radius normalized to the effective radius of M33. The gradient and dispersion in this fit are for regions with R$_{g}$/R$_{e} <$ 2 to isolate the secondary N/O gradient.}
   \label{fig:n_gradients}
\end{figure} 

Nitrogen, which is produced in high-mass stars along with oxygen, has two nucleosynthetic origins: primary nucleosynthesis, in which the ISM is enriched with N via supernovae, and a secondary component where by mass-loss events of intermediate-mass stars releases additional N relative to O and other $\alpha$-elements \citep{henr2000}. As such, N/H and N/O are relevant quantities to understand the total and secondary enrichment of the ISM with N, respectively.

In the top panel of Figure \ref{fig:n_gradients} we plot the N/H abundances against R$_{g,kpc}$, where the gradient and dispersion is provided by the solid line and shaded region, respectively, and the symbol representation is the same as Figure \ref{fig:oh_gradient}. All regions with O/H measurements have significant \nii$\lambda\lambda$6548,6584 detections, hence the number of regions is the same as Figure \ref{fig:oh_gradient}. The 12+log(N/H) gradient in M33 is measured as
\begin{equation}
\mbox{12+log(N/H)}=7.57(\pm0.02) - 0.099(\pm0.007)\times R_{g,kpc}.
\end{equation}
with $\sigma_{int} = 0.044\pm0.014$ dex. The reported intrinsic dispersion in N/H is equivalent to $\sigma_{int}$ about the O/H gradient. This agreement would indicate that the dispersion in O is reflective of the dispersion in other abundant elements in the gas-phase. The total N abundance is related to the O abundance through the ICF(N) $=$ O/O$^+$, which might introduce additional scatter into the N/H trends. However, given the agreement between intrinsic dispersions, this effect does not appear to be large enough to produce a larger $\sigma_{int}$ around the N/H gradient.

The N/H gradient is steeper than the O/H gradient because the ISM has been enriched with secondary N. This is easier to visualize in the bottom panel of Figure \ref{fig:n_gradients}, where we plot the relative abundance of N to O as a function of R$_g$/R$_e$, or galactocentric radius normalized to the effective radius of M33. We briefly mention that the N/O gradient determined from \textsc{LINMIX} when using R$_{kpc}$ is $-0.064\pm0.007$, which is what one expects if the difference between the N/H and O/H gradients is taken.

Instead of fitting a linear gradient to all regions in M33, we focus on the secondary N/O gradient. We follow the methods of \citet{berg2020} and fit the N/O abundances at R$_g$/R$_e <$ 2 with a linear gradient. \citet{berg2020} show that secondary N production is significant below this radius while N is mostly primary in origin beyond R$_g$/R$_e =$ 2.5; they also find that three of the CHAOS galaxies exhibit a common secondary N/O gradient of $-$0.34$\pm$0.06 dex/R$_e$. However, \citet{roge2021} measure a significantly shallower secondary N/O gradient of $-$0.16$\pm$0.04 dex/R$_e$ in the spiral galaxy NGC 2403, which was attributed to this galaxy's lower stellar mass relative to the other CHAOS galaxies. If the stellar mass is a good proxy for the amount of intermediate-mass stars that have enriched the ISM with secondary N, then galaxies with stellar masses similar to NGC 2403 should have similarly shallow secondary N/O gradients.

As shown in the bottom panel of Figure \ref{fig:n_gradients}, the secondary N/O gradient measured in M33 from the 60 \hii regions at R$_g$/R$_e < $ 2 is:
\begin{equation}
\mbox{log(N/O)}=-0.98(\pm0.02) - 0.18(\pm0.02)\times R_{g}/R_{e}.
\end{equation}
with $\sigma_{int} = 0.050\pm0.009$ dex. The secondary N/O gradient in M33 agrees with that of NGC 2403 when considering uncertainties and is significantly shallower than the common gradient measured in NGC 628, M101, and NGC 3184. The stellar mass of M33 is log(M$_\star$/M$_\odot$) $=$ 9.68 \citep{corb2014}, which is very similar to that of NGC 2403 and supports the possible trend of increasing secondary N/O gradient with increasing stellar mass. We note here that the number of galaxies in this sample size is too small to make any robust claims, and that the N/O gradient can be affected by other factors such as galaxy interaction \citep[see NGC 5194]{crox2015,berg2020}. Furthermore, the Milky Way, with stellar mass log(M$_\star$/M$_\odot$) $=$ 10.78 \citep{licq2015}, has a N/O gradient that is significantly shallower \citep[$-$0.05$\pm$0.03 dex/R$_e$,][]{arel2020M} than that observed in M33.

As a final note, the most extended \hii region, M33$-$610$-$1690, appears enhanced in N/O relative to the other extended regions. \nii$\lambda$5755 is not detected in this region, requiring an inferred low-ionization zone temperature for N$^+$/H$^+$ abundances. The linear \te\nii-\te\siii\ relation constructed from the sample of five CHAOS galaxies in \citet{roge2021} is well-sampled and shows low intrinsic dispersion at T$_e <$ 9000 K, but the electron temperatures measured in this region are relatively high: \te\oiii\ $=$ 11,100 K and \te\siii\ $=$ 9,500 K. These temperatures are in an area of parameter space not well sampled by the bulk of the CHAOS \hii regions and, therefore, the relations might not produce realistic values at such high temperatures. Provided that M33$-$610$-$1690 is the only region at R$_g$/R$_e >$ 2.5, we fit the primary N/O plateau considering the five regions at R$_g$/R$_e >$ 2.0 and obtain a weighted-average value of log(N/O)$_{Primary} =$ $-$1.36, which is in good agreement with the primary N/O plateau measured in M101 by \citet{berg2020}.

\begin{figure*}[pt]
\epsscale{1.0}
   \centering
   \includegraphics[width=0.80\textwidth, trim=30 0 30 0,  clip=yes]{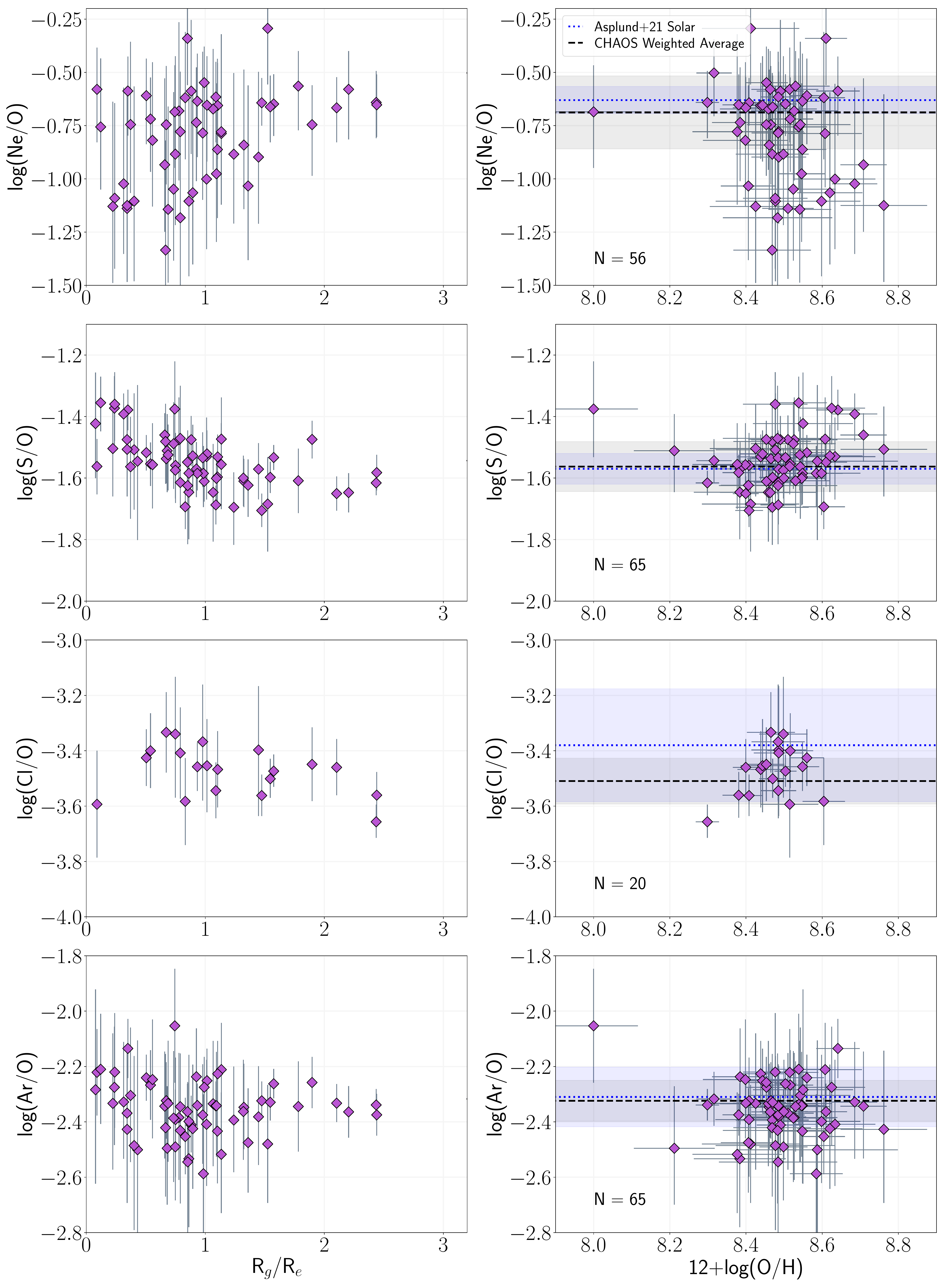}
   \caption{The abundance trends of $\alpha$ elements observed in M33 relative to the oxygen abundance within the region. \textit{Left Column:} The relative abundances plotted against R$_g$/R$_e$. \textit{Right Column:} The relative abundances plotted against 12+log(O/H). The weighted average and standard deviation of the relative abundances are also provided as black dashed lines and gray shaded regions, respectively. The number of \hii regions with the specified relative abundance is provided in the lower left corner. The solar abundance ratios from \citet{aspl2021} are indicated by blue dotted lines and blue shaded regions, respectively. The rows are ordered by increasing atomic number: \textit{First Row:} log(Ne/O); \textit{Second Row:} log(S/O); \textit{Third Row:} log(Cl/O); \textit{Fourth Row:} log(Ar/O).}
   \label{fig:alpha_elements}
\end{figure*} 

\subsection{Neon, Sulfur, Chlorine, and Argon}

The elements Ne, S, and Ar are produced in high-mass stars via the alpha-process, the same mechanism that produces O. The production of Cl is controlled by S and Ar, the former through proton capture and the latter through radioactive decay \citep{clay2003,este2015}. Given their production in the same progenitor stars, the abundance of these four elements should trace the O abundance in an \hii region. In Figure \ref{fig:alpha_elements} we plot the abundance of these elements relative to the oxygen abundance in the regions of M33: the left column plots the relative abundance against R$_g$/R$_e$, the right column plots them against 12+log(O/H), and the rows are ordered in increasing atomic number. In the right column, we provide the weighted average log($\alpha$/O) ratio with its standard deviation as a black dashed line and gray shaded region, respectively. We also provide the solar log($\alpha$/O) ratio and its uncertainty from \citet{aspl2021} as a  blue dotted line and blue shaded region, respectively. Note that we no longer represent the \hii regions by the number of direct temperatures used in the weighted-average ionization zone temperature calculations.

The first row plots the log(Ne/O) relative abundances, which are scattered over $\sim$1 dex. The Ne$^{2+}$/O$^{2+}$ abundances are calculated directly from the emissivities and line intensities using the high-ionization zone temperature because this temperature describes the gas containing both ions. This approach usually produces small uncertainties on the Ne$^{2+}$/O$^{2+}$ abundances because the emissivity ratio of [\ion{Ne}{3}]$\lambda$3868 and \oiii$\lambda$5007 is a weak function of \te. The large uncertainties on log(Ne/O) come from the ICF(Ne), which is relatively uncertain in regions with O$^{2+}$/O $<$ 0.5. This is also exhibited in the additional scatter at log(Ne/O) $<$ $-$1.0, which is where the ICF is rapidly decreasing at O$^{2+}$/O $<$ 0.2. This matches the trends of the \hii region sample used by \citet{amay2021}; their sample of high-ionization blue compact galaxies show a flat log(Ne/O) trend but we do not observe many \hii regions that have this high ionization.

Due to their high uncertainties, the \hii regions with low-ionization
are not weighted heavily in the calculation of the average log(Ne/O) value. The average log(Ne/O) we measure in M33 is log(Ne/O) $=$ $-$0.69$\pm$0.17 from 56 \hii regions. This agrees with the solar value of log(Ne/O) $=$ $-$0.63$\pm$0.06 but note
the large uncertainty. If we only consider \hii regions with O$^{2+}$/O $>$ 0.2 (in other words, excluding the regions where the ICF is most uncertain and is rapidly changing as a function of O$^{2+}$/O), we determine the average log(Ne/O) from 33 regions to be $-$0.63$\pm$0.10. This exact agreement with the solar value and lack of dependence on R$_g$/R$_e$ or 12+log(O/H) reveals that Ne enrichment in M33 is consistent with the trends expected for an element produced predominantly by the $\alpha$ process.

All regions with O abundance have the strong lines of S$^+$ and S$^{2+}$, producing a sample of 65 S/O abundances in M33. The S ICF corrects the ratio of S$^+$ + S$^{2+}$ to O$^+$ + O$^{2+}$ to account for missing S$^{3+}$ in high-ionization \hii regions. All strong lines in this ratio are measured at high S/N and the ICF is well-constrained over a large range of O$^{2+}$/O, resulting in relatively low uncertainties on the log(S/O) in the second row of Figure \ref{fig:alpha_elements}. While some of the inner regions scatter to slightly high S/O, there is no apparent S/O trend with metallicity. The average value measured in M33 is log(S/O) $=$ $-$1.56$\pm$0.08, in excellent agreement with the solar value of $-$1.57$\pm$0.05.

Only 20 regions have significant [\ion{Cl}{3}]$\lambda\lambda$5517,5537 detections, and, in most cases, these emission lines are not detected at high S/N. Furthermore, the uncertainty in the Cl ICF is larger for low-ionization \hii regions which produces the large uncertainties observed in a few objects. The Cl/O relative abundances appear constant as a function of R$_g$/R$_e$, although there is a slight trend of increasing Cl/O as a function of metallicity. This trend is exhibited in the regions with the lowest uncertainty on Cl/O, and \citet{este2020} obtained a similar Cl/O trend in the \hii regions of M101. However, the trend in M101 disappeared depending on the \hii region temperature structure assumed. There are not enough regions in the present sample with O/H $<$ 8.3 or $>$ 8.6 to reliably fit this trend which could, ultimately, be a product of the ICF. The average value of log(Cl/O) in M33 is $-$3.51$\pm$0.08. The solar value reported by \citet{aspl2021}, $-$3.38$\pm$0.20, is relatively uncertain because the Cl abundance is determined through HCl observations in sunspot spectra owing to a lack of Cl features in the Sun's spectrum.

Similar to the strong lines of S, [\ion{Ar}{3}]$\lambda$7135 is observed in all regions with O abundance determinations. Provided this line's high S/N and the potential sky contamination at longer wavelengths, we only use  [\ion{Ar}{3}]$\lambda$7135 in the calculation of Ar$^{2+}$/O$^{2+}$ abundances. There is no trend in Ar/O as a function of R$_g$/R$_e$ or 12+log(O/H), and the average determined from the M33 data, $-$2.32$\pm$0.07, is in very good agreement with the solar value, $-$2.31$\pm$0.11. In summary, the abundances of these four elements in M33 appear to be consistent with the enrichment expected for $\alpha$-process or $\alpha$-process-dependent elements and the relative amount of these elements to O is consistent with the solar ratio. {\it Taken together with the O and N abundances discussed above, M33 is chemically well-mixed and homogeneously enriched from inside-out with no evidence of significant abundance variations at a given radius in the galaxy.}

\vspace{-2mm}
\section{Literature Comparison}
To make the most reliable comparison to the literature direct abundances in M33, we use the line intensities (and their associated uncertainties) and R.A./Decl. centers of each literature \hii region, then recompute the temperatures, abundances, and \hii region radii in a consistent manner. We do this for all \hii regions in M33 from the studies listed in Table \ref{t:m33lit} using the CHAOS reduction steps explained in \S2.2 and \S3.2. We apply the S/N $>$ 3 cutoff for detected lines, which does produce fewer auroral line detections and direct abundances in some studies. For example, one region from \citetalias{croc2006} has a reported \oiii$\lambda$4363 with S/N $<$ 3, resulting in only five regions with \te\oiii. Similarly, \citetalias{lin2017} reported abundances for some regions with non-significant \oiii$\lambda$4363 (S/N $<$ 3), and did not report abundances for regions with significant \nii$\lambda$5755 but non-significant \oiii$\lambda$4363; the net result is that the sample of regions with recalculated direct abundances decreases from 38 to 33.

Most prior studies applied the \te-\te\ relation of \citet{camp1986,garn1992} to obtain the low-ionization zone temperature in the \hii regions. One exception is \citetalias{tori2016}, who used direct \te\oiii\ and \te\nii\ as the high- and low-ionization zone temperatures, respectively, unless \te\oiii\ is undetected. In this case, the high-ionization zone temperature is inferred from \te\nii\ and the empirical relation from \citet{este2009}, which is in good agreement with the relation used by \citet{camp1986} and \citet{garn1992}. The other exception is \citetalias{alex2022}, who took the same approach as \citetalias{tori2016} but did not infer any ionization zone temperature; instead, \citetalias{alex2022} assumed an ionization zone temperature of \te\ = 10$^4$ K when a direct temperature is missing.

We use the weighted-average temperature approach with the \te-\te\ relations of \citet{roge2021} to determine the ionization zone temperature and its uncertainty. For \citetalias{croc2006}, \citetalias{roso2008}, and \citetalias{bres2011}, which do not have the wavelength coverage to measure multiple direct temperatures, this method is simply utilizing the \te-\te\ relations to obtain the electron temperature in the low- and intermediate-ionization zones. For \citetalias{tori2016}, \citetalias{lin2017}, and \citetalias{alex2022}, we use the available \te\oiii\ and \te\nii\ to make a robust estimate of the temperature in each zone. Note that no previous abundance study reports \te\siii\ and we still de-prioritize \te\oii\ and \te\sii, resulting in a weighted average temperature of, at most, one dominant and one inferred \te\ in the low- and high-ionization zone.

\begin{figure*}[pt]
\epsscale{1.0}
   \centering
   \includegraphics[height=0.92\textheight, trim=30 0 30 0,  clip=yes]{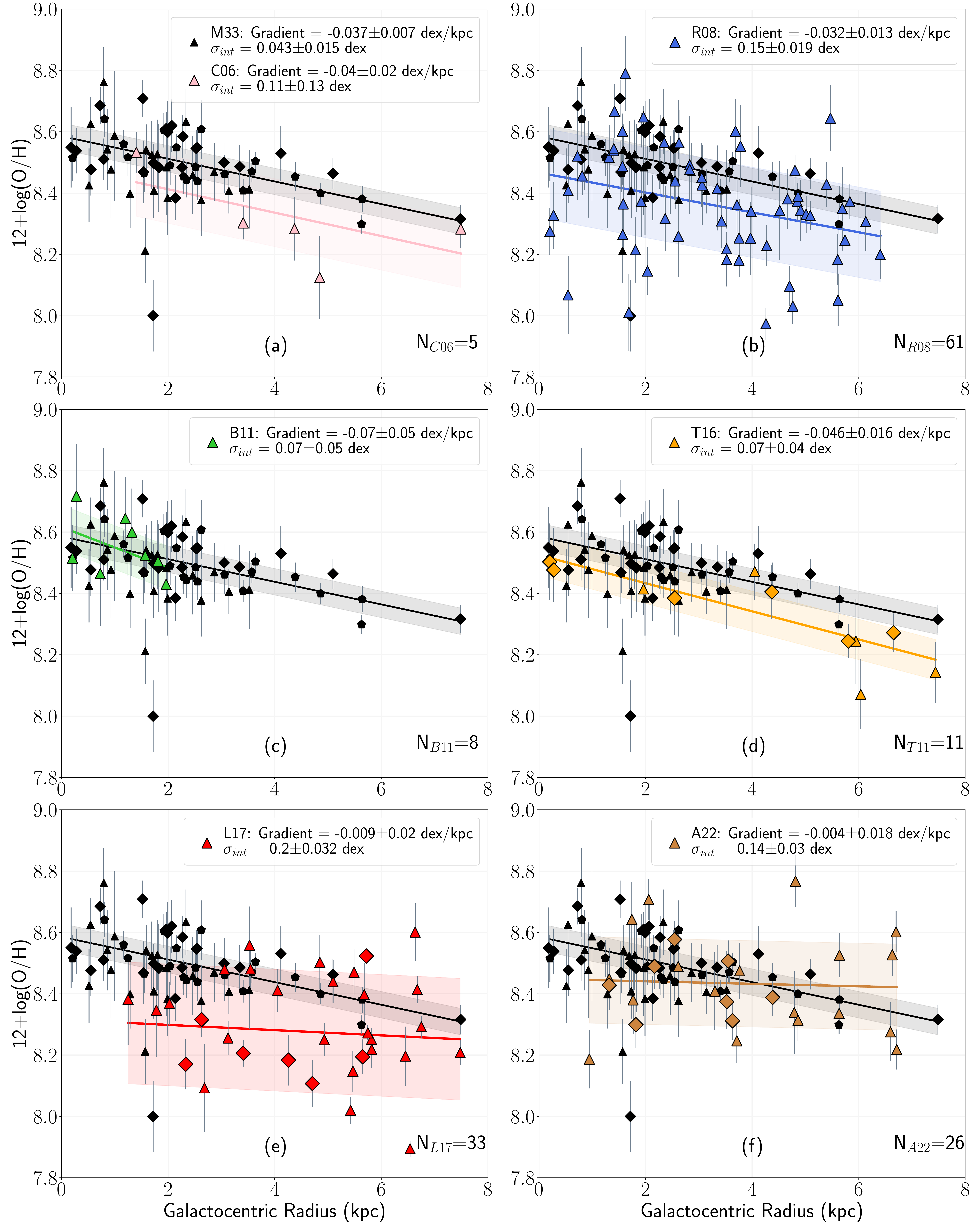}
   \caption{The direct oxygen abundance trends in the M33 \hii regions observed by CHAOS (black points with same representation as Figure \ref{fig:oh_gradient}) and the recalculated literature abundance trends (various colored points with same representation). The gradients and intrinsic dispersions in O/H about the gradient are provided in the legends and are plotted as the solid lines and shaded regions, respectively. The number of literature \hii regions used in the fits are provided in the lower right corner. The panels provide the recalculated abundances of \textit{(a):} \citetalias{croc2006} (pink). \textit{(b):} \citetalias{roso2008} (blue). \textit{(c):} \citetalias{bres2011} (green). \textit{(d):} \citetalias{tori2016} (orange). \textit{(e):} \citetalias{lin2017} (red). \textit{(f):} \citetalias{alex2022} (brown).}
   \label{fig:lit_oh}
\end{figure*} 

Some studies have reported only one of the \oiii\ strong lines; while we use \oiii$\lambda$5007 to obtain the O/H abundance in the CHAOS observations of M33, we use \oiii$\lambda$4959 to recalculate the literature abundances if that is the only available \oiii\ strong line.

In each panel of Figure \ref{fig:lit_oh}, we plot the oxygen abundances of M33 as measured by CHAOS (black points) and the recalculated literature values (various colors depending on the panel), the best-fit \textsc{LINMIX} abundance gradients determined from these regions (solid colored lines) and the intrinsic dispersion in O/H about the gradients (shaded regions around the solid lines). Again, the shapes indicate the number of direct temperatures used in the weighted average ionization zone temperatures with the same representations as Figure \ref{fig:oh_gradient}. In a few instances, namely for \citetalias{croc2006} and \citetalias{bres2011}, the \textsc{LINMIX} gradients/dispersions are not well constrained due to the few direct abundances used in the fitting program. Both \citetalias{croc2006} and \citetalias{bres2011} combined their observations with existing literature data when fitting their gradient and do not report the gradient from their observations alone due to the small number of regions (\citetalias{croc2006}) or the small radial coverage (\citetalias{bres2011}).

For a better constraint on the gradient and dispersion, we instead fit the recalculated \citetalias{croc2006} and \citetalias{bres2011} data with a linear function using the \textsc{Scipy odr} package. This package fits a linear function to the data by minimizing the orthogonal distance of each point to the line of best fit while considering the uncertainties in both dimensions. With the resulting linear fit parameters, we then calculate the intrinsic dispersion about the gradient using \textsc{LINMIX}. In each case, the \textsc{LINMIX} and \textsc{Scipy odr} functions return the same gradient and dispersion, but the uncertainties returned by \textsc{Scipy odr} are more reflective of the data. This technique was repeated for the other literature data compilations, but no difference was measured in the resulting gradients, dispersions, or the uncertainty on either quantity. To remain consistent with the CHAOS approach for M33, the gradients and dispersions in panels (b), (d), (e), and (f) are all calculated using \textsc{LINMIX}.

In general, the recalculated abundances produce gradients that are consistent with the CHAOS-measured gradient. The gradients plotted in panels (a)-(d), which are those reported by studies focusing on the direct and recombination line abundance techniques, agree within uncertainty. Additionally, the intrinsic dispersions about the recalculated \citetalias{croc2006}, \citetalias{bres2011}, and \citetalias{tori2016} abundance gradients agree with the dispersion about the CHAOS M33 gradient, but are not as well constrained due to the smaller number of regions. The dispersion measured about the gradient in panel (b) from the recalculated \citetalias{roso2008} abundances is significantly larger than the dispersion measured by CHAOS and is similar to the dispersion originally reported by \citetalias{roso2008}. The reproduction of the intrinsic dispersion in abundance from the \citetalias{roso2008} line intensity data requires that the atomic data, \te-\te\ relations, and abundance methods are not the source of the large dispersion \citetalias{roso2008} report.

The primarily strong-line studies plotted in panels (e) and (f) produce direct abundance gradients that are significantly shallower than the CHAOS M33 gradient. Unfortunately, the \oiii$\lambda$4363 line in the inner \hii region that \citetalias{lin2017} report (which they term VGV86 II-063) is not at a high enough S/N to be used in abundance determination, producing a smaller radial coverage and a flatter gradient than originally reported. The regions with \te\oiii\ and \te\nii\ tend to scatter low (O/H $<$ 8.2) relative to the CHAOS data. The direct abundances from \citetalias{alex2022} do appear to be well-distributed within the range of abundances expected from the CHAOS data, although there are a few regions at R$_{g} >$ 4 kpc with large O/H that flatten the best-fit gradient. \citetalias{alex2022} label some of these regions as planetary nebula (PNe) based on their position in the classic BPT \citep{bald1981} diagram. Nine of the \hii regions we observe are slightly above the curve that \citet{bald1981} define to designate \hii regions and PNe, but we do not observe large scatter in O/H in these regions. All but one of the eight possible PNe that do have direct abundances in our sample have an oxygen abundance consistent with the gradient and the dispersion about it. As such, we do not remove these regions from our sample when reporting the abundance trends in M33.

\begin{figure}[t!]
\epsscale{1.0}
   \centering
   \includegraphics[width=0.46\textwidth, trim=40 0 40 0,  clip=yes]{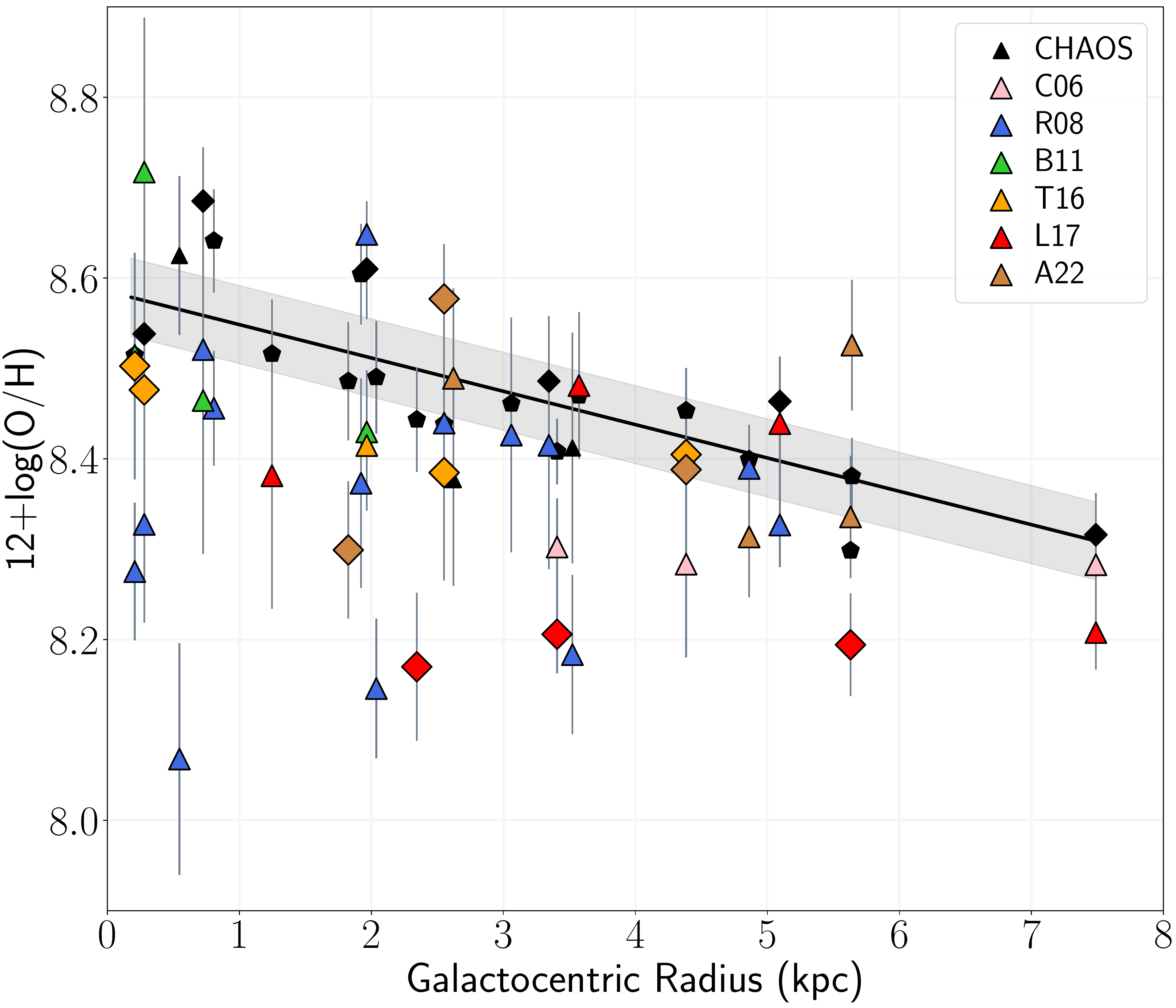}
   \caption{The direct oxygen abundances in the common \hii regions of M33 measured by CHAOS (black points) and the literature (other colors, same as Figure \ref{fig:lit_oh}). The best-fit CHAOS gradient and intrinsic dispersion about the gradient are provided as the solid black line and shaded gray region, respectively. The shape of the points represents the number of direct temperatures used to calculate the abundances, same as Figure \ref{fig:oh_gradient}. \hii regions which contain CHAOS abundances but are missing literature abundances are not plotted.}
   \label{fig:lit_oh_gradient}
\end{figure} 

\begin{figure*}[!t] 
\epsscale{1}
   \centering
   \includegraphics[width=0.95\textwidth, trim=30 0 30 0,  clip=yes]{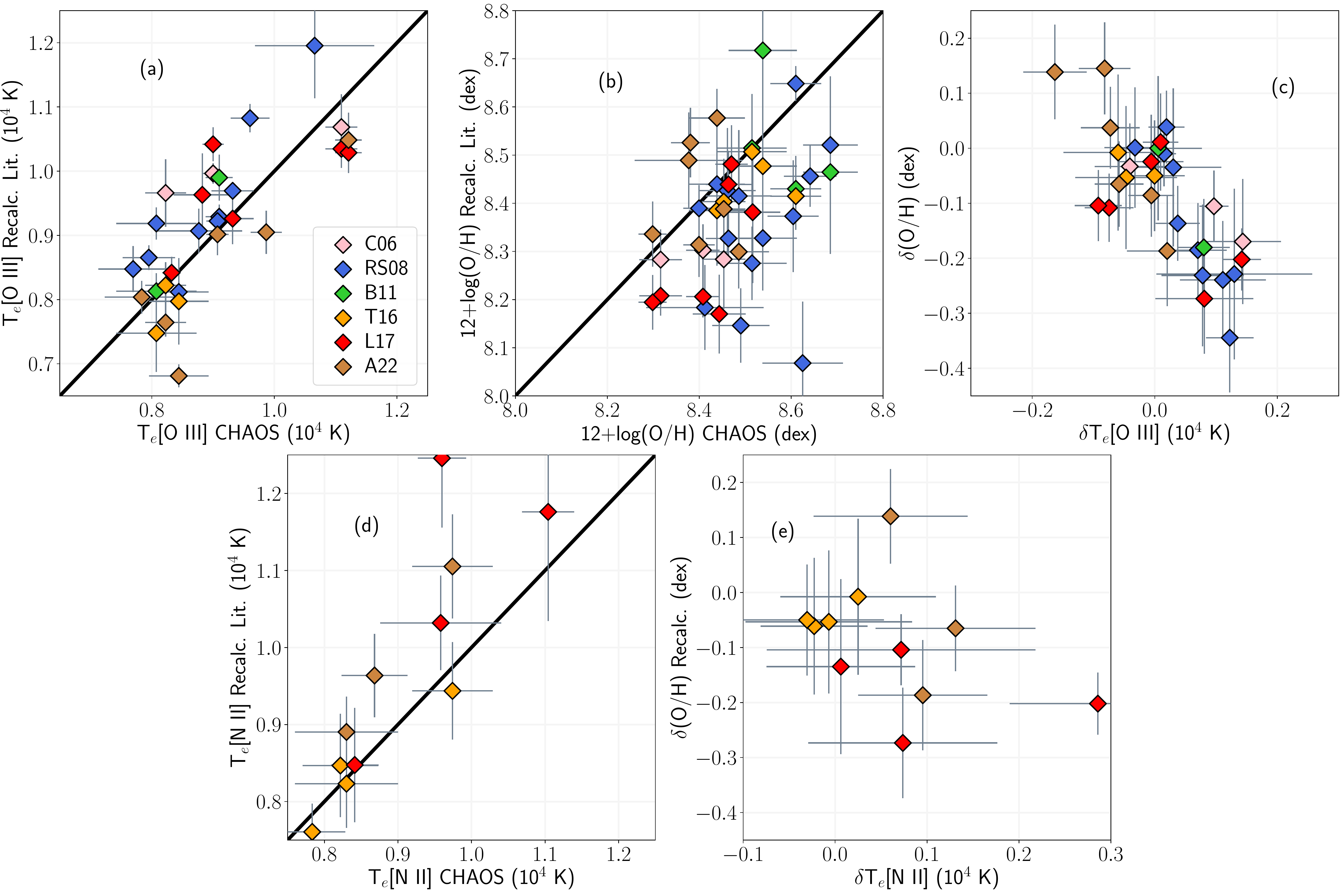}
   \caption{A comparison of the recalculated \te\oiii, \te\nii, and 12+log(O/H) in the common \hii regions observed by the literature studies and CHAOS. Each point is color-coded based on the literature study (see legend). \textit{(a):} Recalculated \te\oiii\ from the literature study vs. the CHAOS \te\oiii\ measured in the same region. The black solid line represents a one-to-one relation. \textit{(b):} Recalculated 12+log(O/H) from the literature study vs. the CHAOS 12+log(O/H). \textit{(c):} $\delta$(O/H), defined as the difference between the CHAOS and literature oxygen abundances in the same region, vs. $\delta$\te\oiii, the difference between the CHAOS and literature \te\oiii. \textit{(d)}: Recalculated \te\nii\ from the literature study vs. the CHAOS \te\nii. \textit{(e)}: $\delta$(O/H) vs. $\delta$\te\nii.}
   \label{fig:fullcompfig}
\end{figure*} 

Similar regions are targeted in the literature and the present study (see Table \ref{t:m33Obs}), allowing for a more direct comparison of the abundances measured in each region. In Figure \ref{fig:lit_oh_gradient}, we plot the CHAOS direct oxygen abundances and gradient in black and the direct oxygen abundances measured in the same \hii regions by the literature studies (same colors as Figure \ref{fig:lit_oh}). In general, the CHAOS oxygen abundances are consistent or larger than the literature abundances for the same \hii regions. Additionally, we report a large range of abundances for some of the most well-studied \hii regions in M33. For example, three literature studies (\citetalias{roso2008}, \citetalias{bres2011}, \citetalias{tori2016}) observe M33$-$36$-$52, a bright, central \hii region that is crucial for constraining the O/H gradient at low radii. The recalculated oxygen abundances for these literature studies and for CHAOS are (O/H)$_{R08} =$ 8.33$\pm$0.11, (O/H)$_{B11} =$ 8.72$\pm$0.17, (O/H)$_{T16} =$ 8.48$\pm$0.10, and (O/H)$_{CHAOS} =$ 8.54$\pm$0.08, which represent a potential range of $\sim$0.4 dex in abundance between the various studies. We note that the studies with broad wavelength coverage, relatively high spectral resolution, and more than one \te-sensitive line detected/applied (\citetalias{tori2016} and this study) agree within uncertainty.

In Figure \ref{fig:fullcompfig}, we attempt to uncover the source of the difference in abundance within these common \hii regions and, potentially, the large scatter in the previous abundance studies. In panel (a), we plot the recalculated \te\oiii\ against the CHAOS measured \te\oiii\ in these regions. Overall, the trend in panel (a) reveals that the literature and CHAOS \te\oiii\ track each other relatively well, although there is scatter about the line of unity. This is the case for the majority of the temperatures from \citetalias{roso2008}, where all but one of the common regions has a higher recalculated \te\oiii\ when compared to the CHAOS temperatures. Defining the difference between the CHAOS measured and literature \te\oiii\ as $\delta$\te\oiii, the average $\delta$\te\oiii\ for \citetalias{roso2008} is $>$ 600 K. There are an additional four common \hii regions that contain a recalculated \te\oiii\ from the \citetalias{roso2008} line intensities but in which we fail to detect the \oiii$\lambda$4363 auroral line. This is similar to what \citetalias{bres2011} finds in their sample of 25 central \hii regions: \oiii$\lambda$4363 is undetected in 17 of the \hii regions, four of which \citetalias{roso2008} report a detection of \oiii$\lambda$4363.

Panel (b) plots the recalculated literature oxygen abundances in the common regions against the abundance we report for these regions. The CHAOS abundances are calculated using the available \te\oiii, \te\siii, and \te\nii\ data, which may result in more regions for comparison relative to panel (a). As Figure \ref{fig:lit_oh_gradient} makes evident, the recalculated abundances for the literature studies \citetalias{roso2008} and \citetalias{lin2017} are scattered lower than the CHAOS abundances. The average $\delta$(O/H), or the difference in direct oxygen abundance between the CHAOS and the literature, is $-$0.12 dex and $-$0.17 dex for \citetalias{roso2008} and \citetalias{lin2017}, respectively. \citetalias{alex2022} recalculated abundances are scattered above and below the line of unity, resulting in an average $\delta$(O/H) of $-$0.02 dex. There are too few overlap regions to make a thorough comparison for \citetalias{croc2006} and \citetalias{bres2011}, but the abundances in the common regions observed by \citetalias{tori2016} are in excellent agreement with the CHAOS abundances (average $\delta$(O/H) $=$ $-$0.05 dex).

The source of the scatter in panel (b) is explored in panel (c), in which we plot $\delta$(O/H) against $\delta$\te\oiii. The anti-correlation between $\delta$(O/H) and $\delta$\te\oiii\ is clear: the scatter to low oxygen abundances relative to the CHAOS abundances is related to significantly higher \te\oiii\ in these regions. This is supported by the two regions from \citetalias{alex2022} with $\delta$\te\oiii\ $<$ 0 and $\delta$(O/H) $>$ 0, and is consistent with the expected trend that electron temperature is anti-correlated with abundance. For three of the literature studies, \te\oiii\ ultimately controls the entirety of the oxygen abundance: O$^{2+}$ abundance directly from \te\oiii\ and the O$^+$ abundance through the inferred low-ionization zone temperature. Significantly different \oiii\ temperatures should produce noticeably different oxygen abundances, which indicates that the source of the scatter to low abundance is higher \te\oiii/larger \oiii$\lambda$4363/\oiii$\lambda$5007 intensity ratios in the literature. There is a cluster of regions around $\delta$(O/H) $=$ 0 and $\delta$\te\oiii\ $=$ 0 from regions with very similar temperatures and abundances; an exact agreement in temperature may still produce scatter around this location due to slight differences in \oiii\ strong-line intensities and the applied low- and high-ionization zone temperatures (the latter not necessarily equivalent to \te\oiii\ due to the weighted averages).

Panel (d) is the same as panel (a) except the recalculated and CHAOS \te\nii\ are plotted. There is very good agreement between the recalculated \te\nii\ of \citetalias{tori2016} and those reported in this work. However, the agreement becomes less clear when considering the recalculated temperatures of \citetalias{lin2017} and \citetalias{alex2022}, where both studies produce slightly larger \te\nii\ than measured from the CHAOS observations. These higher \te\nii\ could be responsible for the flattening of the O/H gradient and the scatter to low O/H observed in both of these studies, although the uncertainties on these \te\nii\ are rather large and, therefore, are not prioritized as much as \te\oiii\ in the weighted average ionization zone temperatures.

\begin{figure*}[!ht] 
\epsscale{1}
   \centering
   \includegraphics[width=0.95\textwidth, trim=30 0 30 0,  clip=yes]{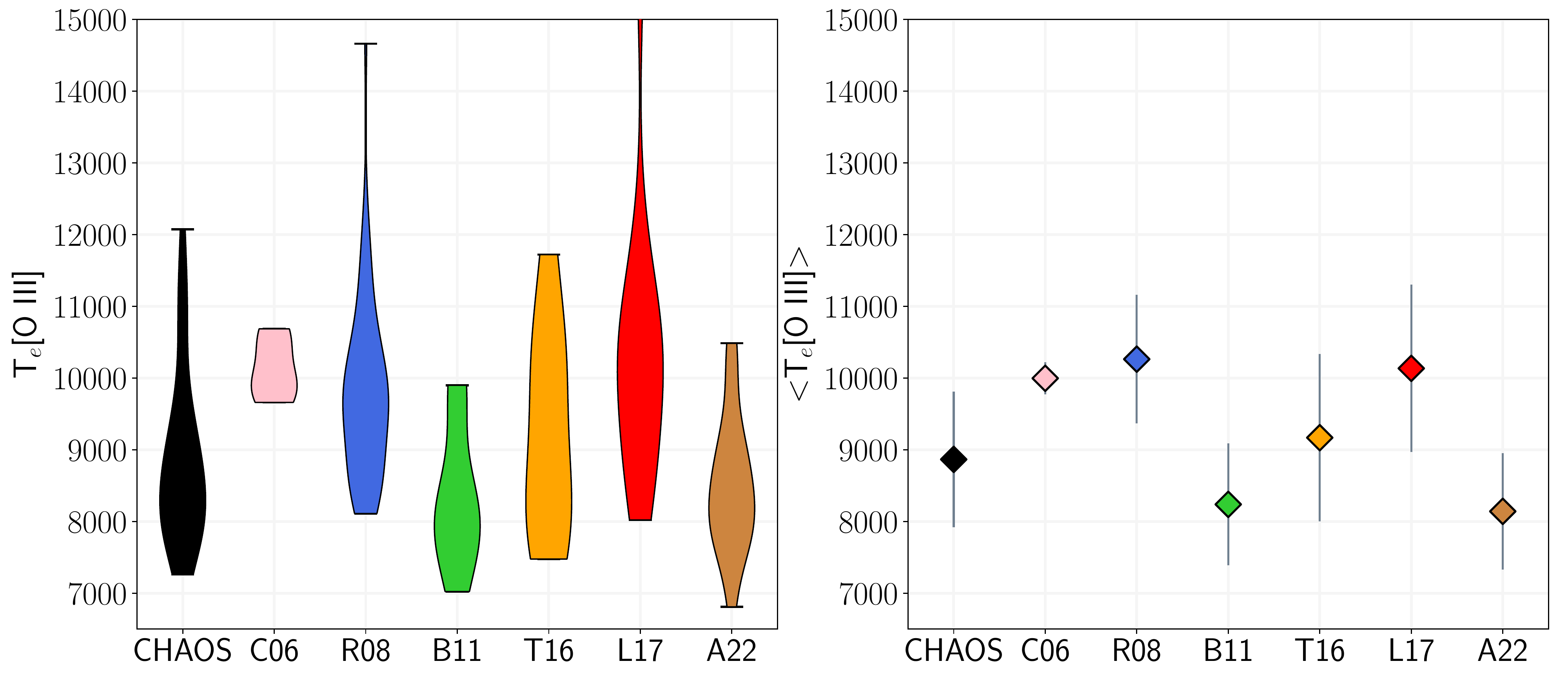}
   \caption{The distribution of \te\oiii\ in M33 measured by CHAOS and recalculated from the literature studies. \textit{Left Panel:} A violin plot of the \te\oiii\ measured from the different studies. The minimum, maximum, and number density of \te\oiii\ are plotted as the base, peak, and width of each violin, respectively. Some of the violin profiles are affected by the relatively few recalculated \te\oiii\ (for example, \citetalias{croc2006} and \citetalias{tori2016}). \textit{Right Panel:} The average \te\oiii\ and standard deviation measured from the distribution of \te\oiii\ in each sample.}
   \label{fig:to3_violin}
\end{figure*} 

The anti-correlation between $\delta$(O/H) and $\delta$\te\nii\ is not as clear in panel (e): while the common regions of \citetalias{tori2016} are still clustered around the origin, the low number of regions and the larger uncertainties on $\delta$\te\nii\ do not make as convincing a trend. As previously mentioned, the large \te\nii\ uncertainties could produce a low-ionization zone temperature that is weighted more to the inferred temperature from a well-measured \te\oiii; if this is the case, then the uncertain \te\nii\ has little impact on the final O/H and, therefore, may produce no anti-correlation with O/H.

While we have focused on the common \hii regions from the literature and our sample, these high \oiii\ temperatures are likely the source of the scatter observed in the other literature \hii regions. In Figure \ref{fig:to3_violin}, we plot the distribution of \te\oiii\ observed in CHAOS and the literature studies. The violin plot in the left panel shows the minimum, maximum, and number density of \te\oiii\ (as minimum, maximum, and width of the violin, respectively) measured in each sample, while the right panel plots each sample's average \te\oiii\ and its standard deviation. We note that 1. We do not expect all \te\oiii\ to be in exact agreement because the electron temperature is a function of metallicity and each study targets different \hii regions, and 2. That the extreme temperatures affect both the height of the violins and the standard deviation of the average \te\oiii. The second point is what produces the large standard deviations observed in the right panel, resulting in an agreement between all average \te\oiii. However, the left panel reveals that the density of \te\oiii\ from CHAOS, \citetalias{bres2011}, and \citetalias{alex2022} all peak at roughly the same temperature. Not only does the density of \oiii\ temperatures from \citetalias{roso2008} and \citetalias{lin2017} peak at higher \te\oiii, but the maximum recalculated \te\oiii\ from these samples is larger than observed in any other literature study. Some studies have relatively few recalculated \te\oiii, which impacts their appearance on the violin plot (5 for \citetalias{croc2006}, 8 for \citetalias{bres2011}, and 9 for \citetalias{tori2016}). Of particular note is \citetalias{tori2016}, which shows a relatively constant \te\oiii\ density from the minimum to the peak \te\oiii. This is likely due to the radial and abundance range covered by this sample: \citetalias{tori2016} measured \oiii$\lambda$4363 in nine \hii regions ranging from the center to the outskirts of M33, producing a broad span of \te\oiii\ due to the change in O/H but not well-sampled enough to produce a clear peak in the \te\oiii\ distribution. Overall, \citetalias{tori2016} measure a range of \oiii\ temperatures similar to ours (see also Figure \ref{fig:fullcompfig}).

\subsection{A Discussion on the Temperature Dispersion in M33}

The remaining question is, then, why are the \te\oiii\ from some of the literature studies offset to larger values than we measure in M33? \te\oiii\ is dependent on \oiii$\lambda$4363 such that a stronger auroral line produces a larger auroral-to-nebular line ratio which, in turn, results in a higher \te\oiii. Perhaps the simplest explanation is that noise in the continuum could be interpreted for auroral line emission. If the noise in the continuum is not reflected in the emission line uncertainties and is assumed to be physical, then the reported \oiii\ auroral line intensities would be biased to larger temperatures resulting in underestimated \hii region oxygen abundances.

Alternatively, \oiii$\lambda$4363 can be contaminated by other emission features. One possible source of contamination is [\ion{Fe}{2}]$\lambda$4360, which would bias a single fit to \oiii$\lambda$4363 high if unaccounted for \citep{curt2017}. This is particularly the case if [\ion{Fe}{2}]$\lambda$4360 is comparable to a \oiii$\lambda$4363 line that is, on its own, not significant, which might be the situation for high-metallicity \hii regions. In this scenario, the auroral line goes from being undetected (with no corresponding direct temperature) to being significantly detected and producing an unrealistically large temperature and low O$^{2+}$ abundance. Provided that many of the studies examined in this work do not have the spectral resolution to fully deblend [\ion{Fe}{2}]$\lambda$4360 and \oiii$\lambda$4363, and that these studies did not account for [\ion{Fe}{2}] contamination, this is a potential explanation for the scatter towards high \te\oiii/low O/H.

The degree of significant [\ion{Fe}{2}] contamination in M33 that we report in the CHAOS sample is present, but minimal in that few regions have [\ion{Fe}{2}]$\lambda$4288 with S/N $>$ 3. Of the nearly 100 regions we observe, only four\footnote{M33$-$224$-$346, which is characterized by emission from the LBV M33C-7256, also has extremely intense [\ion{Fe}{2}]$\lambda$4288 and [\ion{Fe}{2}]$\lambda$4360, but it is not used in \te\ or abundance analysis and, therefore, not counted.} have significant detections of [\ion{Fe}{2}]$\lambda$4288 and, therefore, significant [\ion{Fe}{2}]$\lambda$4360 emission that must be accounted for when fitting \oiii$\lambda$4363. As described in \S3.1, we subtract the inferred [\ion{Fe}{2}]$\lambda$4360 intensity from the intensity of \oiii$\lambda$4363 to account for [\ion{Fe}{2}] contamination. This method ensures that [\ion{Fe}{2}]$\lambda$4360 is completely removed from \oiii$\lambda$4363 and utilizes a line that is in an area of the continuum that is easy to fit. MODS has a resolution of R $\approx$ 2000 in the blue, which allows for the fit to [\ion{Fe}{2}]$\lambda$4360 in cases where the profile of the \oiii$\lambda$4363 line is asymmetric. We have done these fits for all \hii regions with asymmetric \oiii$\lambda$4363 and for regions with 2 $<$ S/N([\ion{Fe}{2}]$\lambda$4288) $<$ 3 to verify that the profile of \oiii$\lambda$4363 can be trusted.

\begin{figure}[t]
\epsscale{1.0}
   \centering
   \includegraphics[width=0.45\textwidth, trim=40 0 40 0,  clip=yes]{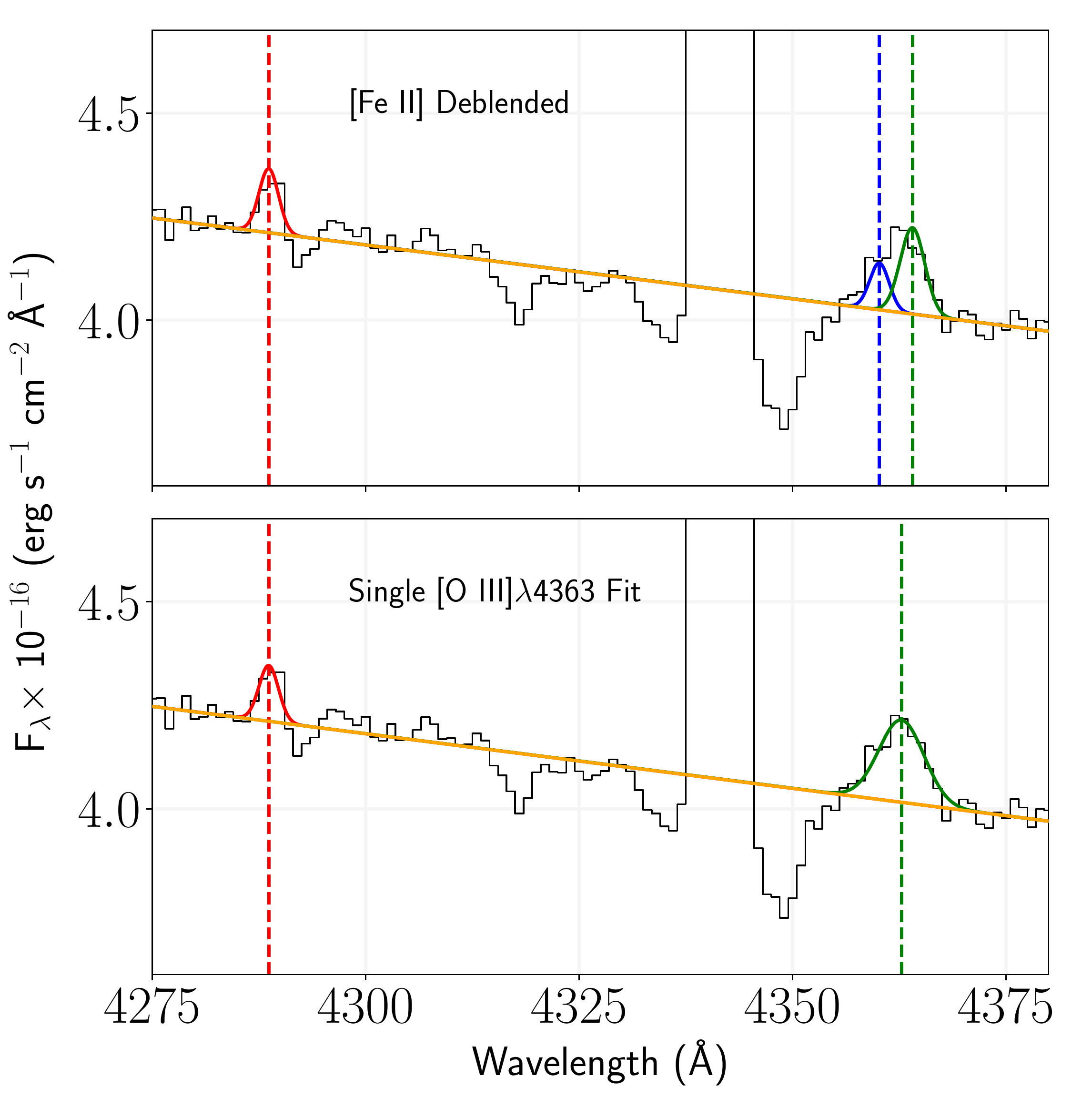}
   \caption{The spectrum of M33$-$35$-$385 near H$\gamma$. The lines [\ion{Fe}{2}]$\lambda$4288, [\ion{Fe}{2}]$\lambda$4360, and \oiii$\lambda$4363 can be made out in this wavelength range. \textit{Top Panel}: A fit to the three emission lines while using the \oiii\ strong lines to constrain the Gaussian FWHM of \oiii$\lambda$4363 and the same FWHM for each [\ion{Fe}{2}] line. Each line fit is provided as a colored Gaussian while the linear continuum is plotted as an orange line. [\ion{Fe}{2}]$\lambda$4360 and \oiii$\lambda$4363 can be deblended and produces a non-detection of \oiii$\lambda$4363. \textit{Bottom Panel}: A fit to [\ion{Fe}{2}]$\lambda$4288 and a line at 4361.5 \AA\ to simulate a single Gaussian fit to \oiii$\lambda$4363. The Gaussian for \oiii$\lambda$4363 is clearly too intense and too broad relative to the Gaussian with FWHM set by the strong lines, but the intensity is large enough to produce an erroneous line detection.}
   \label{fig:fe_contam}
\end{figure} 

We plot an example of this in Figure \ref{fig:fe_contam}, which is the spectrum of the \hii region M33$-$35$-$385 focused around H$\gamma$ and \oiii$\lambda$4363. From both panels, there is clear [\ion{Fe}{2}]$\lambda$4288 emission and emission at 4363 \AA, but the S/N of [\ion{Fe}{2}]$\lambda$4288 is less than 3. In the top panel we fit [\ion{Fe}{2}]$\lambda$4288, [\ion{Fe}{2}]$\lambda$4360, and \oiii$\lambda$4363. To fit the three lines in the top panel, we first fit the \oiii\ strong lines to obtain the FWHM of the Gaussians in km/s, then use this to obtain the FWHM of the \oiii$\lambda$4363 Gaussian. We then assume that the [\ion{Fe}{2}] lines have the same FWHM and that their flux ratio must be equivalent to the emissivity ratio of the two transitions. The fit to [\ion{Fe}{2}]$\lambda$4360 and \oiii$\lambda$4363 captures the asymmetric profile and one can clearly see the contamination from [\ion{Fe}{2}]$\lambda$4360 despite the fact that [\ion{Fe}{2}]$\lambda$4288 is measured at 2 $<$ S/N([\ion{Fe}{2}]$\lambda$4288) $<$ 3. The bottom panel plots the fit to [\ion{Fe}{2}]$\lambda$4288 and \oiii$\lambda$4363 assuming that there is a single line at 4361.5 \AA\ and allowing for the Gaussian FWHM of each line to vary freely. The intensity and width of the line at 4361.5 \AA\ is clearly too large relative to the deblended fit in the top panel, which would be the scenario if the resolution of MODS was insufficient to deblend [\ion{Fe}{2}]$\lambda$4360 and \oiii$\lambda$4363. The fit in the top panel produces a non-detection in \oiii$\lambda$4363 while the bottom panel is at sufficiently high signal to produce a detection, which would bias the \te\oiii\ and ionization-zone temperatures in this region to unphysically large values.

We propose an additional explanation for the high \te\oiii\ in some of the literature studies: contamination from the night sky. The redshift of M33, z $= -$0.000597, is particularly troublesome for direct abundance studies that rely solely on \oiii$\lambda$4363. At this redshift, \oiii$\lambda$4363 is blueshifted from its theoretical wavelength to $\lambda$4360.60 \AA\ \citep[with individual \hii regions shifted up to $\pm$ $\sim$ 2\AA\ due to the rotation of the galaxy;][]{koch2018} which is in close proximity to the \ion{Hg}{1} night sky line at $\lambda$4358.34 \AA. If there is prominent \ion{Hg}{1} night sky emission and the spectral resolution of the detector is not sufficient, these two lines could be partially or totally blended. \ion{Hg}{1} $\lambda$4358 is observed at the LBT, but the resolution is sufficient that the night sky line can be cleanly subtracted in most cases. As described in \S3.1, in cases where the profile of \oiii$\lambda$4363 is asymmetric or there is evidence of poor subtraction of the \ion{Hg}{1} line, we fit the sky residuals to the blue side of \oiii$\lambda$4363 and exclude this emission from the flux of \oiii$\lambda$4363. This approach was originally taken for potential significant [\ion{Fe}{2}] contamination, but it works for a  \ion{Hg}{1} $\lambda$4358 undersubtraction artifact because this noise is shifted to 4360.9 \AA\ when the spectrum is redshift corrected.

\ion{Hg}{1} contamination of \oiii$\lambda$4363 will typically produce \oiii\ temperatures that are biased higher than the actual \oiii\ temperature in the region. Consider a spectrum with insufficient resolution to completely deblend \ion{Hg}{1} $\lambda$4358 and \oiii$\lambda$4363 in M33. There are two scenarios of contamination, one in which the \ion{Hg}{1} line is oversubtracted and another in which the line is undersubtracted. In the first case, the oversubtraction of \ion{Hg}{1} clips the \oiii\ auroral line and produces negative residuals on the blue side of the otherwise Gaussian profile. In cases of extreme contamination, the loss in \oiii$\lambda$4363 flux will produce a S/N $<$ 3 and cause the auroral line to go undetected. In the second scenario, an undersubtraction still produces a non-Gaussian profile but increases the flux in the fit for \oiii$\lambda$4363. When contamination is large, the additional flux could cause an otherwise undetected \oiii$\lambda$4363 to be fit at a S/N $>$ 3, but even small amounts of contamination will bias the \oiii\ flux/temperature high. In this way, \ion{Hg}{1} contamination removes all 4363 when the oversubtraction is significant but increases 4363 when there is any undersubtraction in the line, producing a net trend to higher \oiii\ flux.

\begin{figure}[t]
\epsscale{1.0}
   \centering
   \includegraphics[width=0.45\textwidth, trim=40 10 40 10,  clip=yes]{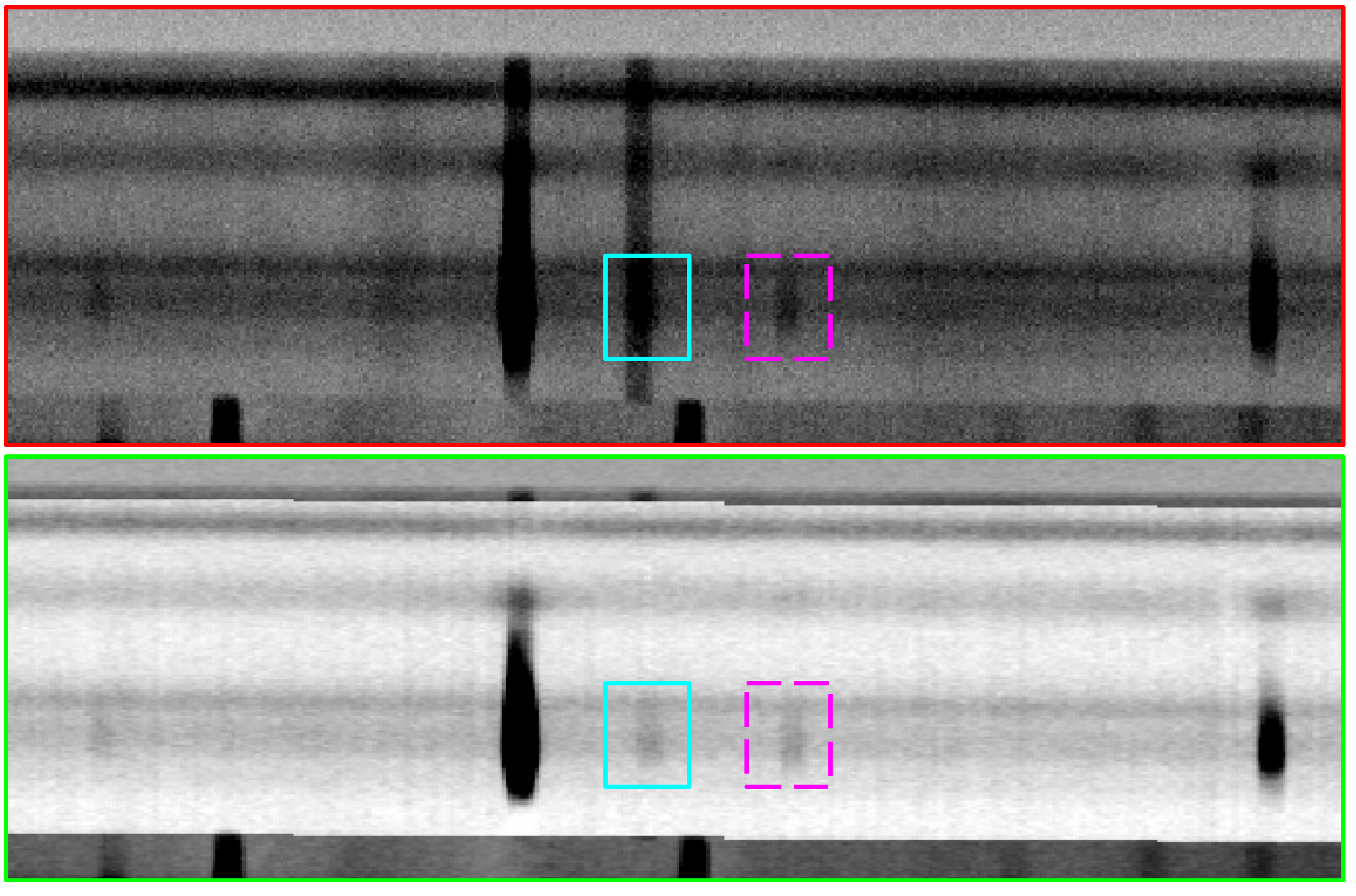}
   \caption{A portion of the MODS 2D spectrum of M33+29+261 focused on H$\gamma$, \ion{Hg}{1} $\lambda$4358, \oiii$\lambda$4363, \ion{He}{1} $\lambda$4388, and \ion{He}{1} $\lambda$4471. \textit{Top Panel}: The 2D spectrum before sky subtraction. The location of \oiii$\lambda$4363 is highlighted in the cyan box and the line can be seen on the red edge of the \ion{Hg}{1} $\lambda$4358 emission. With lower spectral resolution, \oiii$\lambda$4363 may be entirely blended with \ion{Hg}{1} $\lambda$4358. The location of \ion{He}{1} $\lambda$4388 is highlighted with the dashed magenta box. \textit{Bottom Panel}: The 2D spectrum after sky subtraction. \ion{Hg}{1} $\lambda$4358 is cleanly subtracted and the \oiii$\lambda$4363 emission is present within the cyan box. The 2D profile of \oiii$\lambda$4363 is comparable to the \ion{He}{1} $\lambda$4388 feature, the latter of which is not susceptible to \ion{Hg}{1} contamination.}
   \label{fig:2dsky}
\end{figure} 

The potential \ion{Hg}{1} contamination in the literature spectra of M33 is dependent on: 1. The resolution of the detector, and 2. Whether or not \ion{Hg}{1} emission is present in the sky at the telescope location. For instance, the raw data obtained from the GTC used by \citetalias{tori2016} show evidence of \ion{Hg}{1} $\lambda$4358, but the resolution of the R2500U grism is sufficient enough to clearly separate this line from \oiii$\lambda$4363. An example of a MODS 2D spectrum obtained from the multi-object field observation of M33+29+261 is provided in Figure \ref{fig:2dsky}. The top panel is the spectrum before sky subtraction, while the bottom is the spectrum after sky subtraction (the limits are unchanged between the panels). We focus on the portion of the spectrum from H$\gamma$, the intense line to the left of the cyan box, to \ion{He}{1} $\lambda$4471. The locations of \oiii$\lambda$4363 and \ion{He}{1} $\lambda$4388 are highlighted by the solid cyan box and dashed magenta box, respectively, in both panels. In the top panel, the \ion{Hg}{1} sky line spans the entire slit and is close in proximity to \oiii$\lambda$4363, but it is possible to make out the red edge of the auroral line. In the bottom panel, \ion{Hg}{1} $\lambda$4358 line is cleanly subtracted and the \oiii\ auroral line emission is still detectable. Furthermore, the 2D profile of \oiii$\lambda$4363 is comparable to the \ion{He}{1} $\lambda$4388 profile in the dashed magenta box. This line is not affected by \ion{Hg}{1} or night sky contamination, therefore the similar profile of the \oiii\ auroral line would indicate negligible sky contamination in the MODS spectrum.

Detectors with insufficient spectral resolution, however, could be subject to the night sky or [\ion{Fe}{2}] contamination. This could be a potential explanation for the scatter to high \te\oiii\ observed in the recalculated \citetalias{lin2017} temperatures: the resolution of $\sim$6 \AA\ of Hectospec at the MMT might be insufficient to deblend the \oiii\ auroral line and night sky line, in which case the contamination will bias the \oiii\ flux and \te\ high. The same is true for \citetalias{roso2008}, who obtained their raw spectra using LRIS on Keck I, but the degree to which \ion{Hg}{1} $\lambda$4358 is present at Mauna Kea would favor [\ion{Fe}{2}] contamination. The agreement between the CHAOS and \citetalias{tori2016} \oiii\ and \nii\ temperatures, the latter of which are free from these sources of contamination, indicate that high-S/N observations with multiple \te-sensitive auroral lines produce the same abundance gradient and similarly low scatter in abundance. This highlights the importance of high spectral resolution to distinguish various emission lines, and large wavelength coverage to properly measure the temperatures in different ionization zones in an \hii region and to avoid a reliance on a single, potentially contaminated \te\ measurement.

\begin{figure}[t]
\epsscale{1.0}
   \centering
   \includegraphics[width=0.45\textwidth, trim=40 0 40 0,  clip=yes]{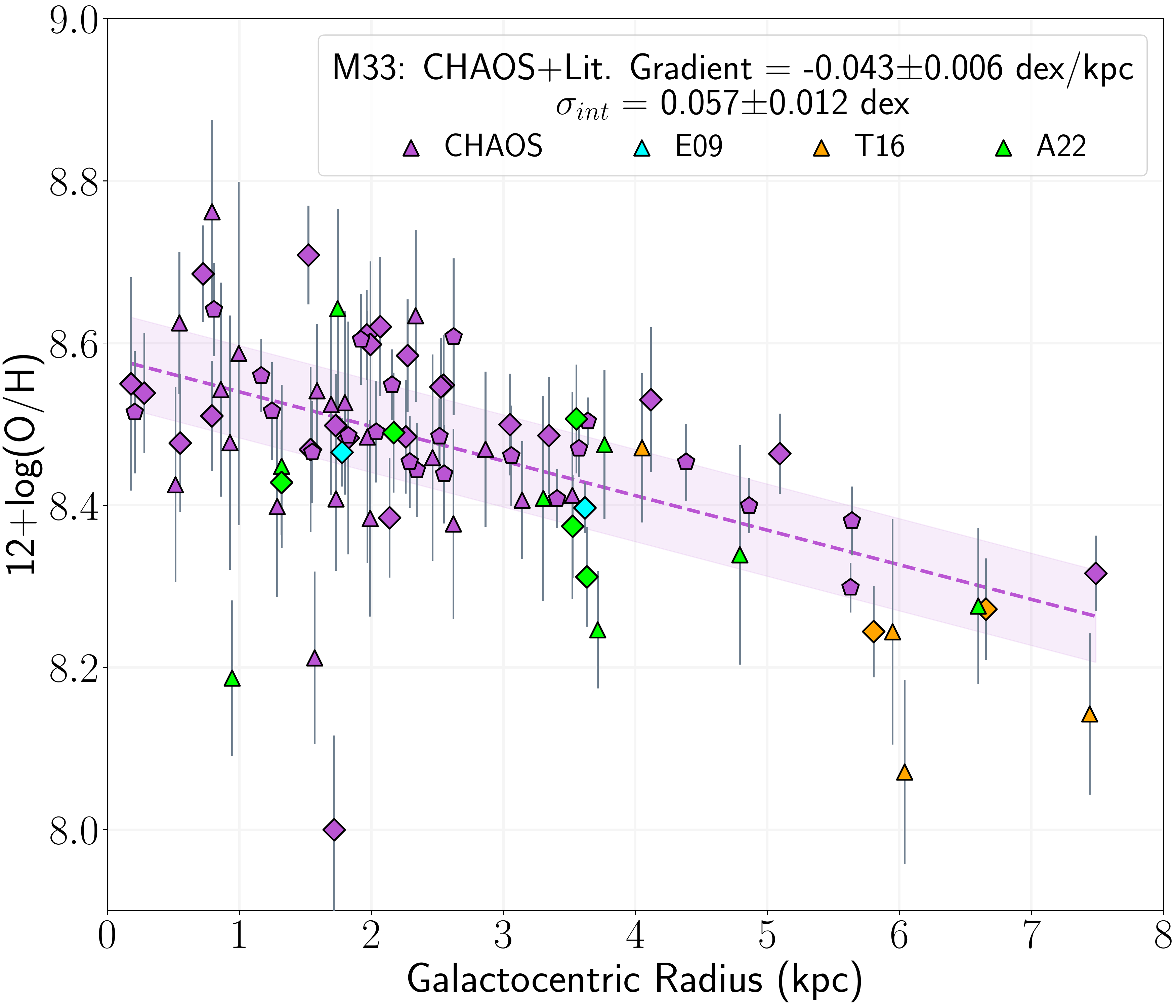}
   \caption{The oxygen abundance gradient in M33 as measured from CHAOS (65 regions, purple), \citetalias{tori2016} (6 regions, orange), \citetalias{alex2022} (13 regions, green), and \citet{este2009} (2 regions, cyan). The best-fit gradient (dashed purple line) and intrinsic dispersion in O/H (shaded purple area) are provided in the legend. The point representation is the same as Figure \ref{fig:oh_gradient}.}
   \label{fig:oh_comp_grad}
\end{figure} 

We believe the homogeneity of the CHAOS sample makes for the most reliable measurement of properties such as the intrinsic dispersion about the abundance gradient. Nonetheless, we determine the impact on the oxygen abundance gradient and dispersion when we combine the CHAOS abundances with those from the literature with similar wavelength coverage, spectral resolution, and direct temperature measurements. These studies include \citetalias{tori2016}, \citetalias{alex2022}, and the abundances of NGC 604 and NGC 595 from \citet{este2009}, which have been recalculated in the same manner as the other literature samples. \citet{este2009} used the Keck High Resolution Echelle Spectrometer (HIRES), which has a spectral resolution of R$\approx$23,000 (for their configuration) and is sufficient to deblend any potential \oiii$\lambda$4363 contaminators. In constructing the combined M33 sample, we exclude the PNe objects from \citetalias{alex2022} due to the unexplained scatter to high O/H that we do not observe. In cases of overlap regions, we prioritize our homogeneous sample, followed by \citetalias{tori2016}, \citet{este2009}, and finally \citetalias{alex2022}. The sample of 86 \hii regions is plotted as a function of radius (in kpc) in Figure \ref{fig:oh_comp_grad}, and the resulting best-fit O/H gradient is $-$0.043$\pm$0.006 dex/kpc with intrinsic dispersion $\sigma_{int} =$ 0.057$\pm$0.012 dex. Despite the increase in the slope of the composite gradient due to the inclusion of the low abundances observed in the outer \hii regions by \citetalias{tori2016} (see panel d in Figure \ref{fig:lit_oh}), both the gradient and intrinsic dispersion agree with the values obtained from the homogeneous CHAOS sample.

\subsection{The Abundance Dispersion in Local Spirals}

The dispersion in O/H measured about the abundance gradient in a spiral galaxy is related to the presence/absence of mixing mechanisms within the galaxy. Various mechanisms affect gas motion and mixing on different time and spatial scales \citep[see discussion in][]{roy1995}. The observation of low dispersion about an abundance gradient is indicative of effective mixing mechanisms that distribute the heavy elements after local enrichment from stellar nucleosynthesis. On the contrary, large abundance variations indicate that local enrichment is dominant, i.e. that the mechanisms acting to homogenize the ISM are inefficient or not present. From the large sample of regions in M33, the dispersion in O/H is relatively low/consistent with the observational uncertainties on the most well-measured abundances. The subsample of regions with the direct \te\ from \oiii, \siii, and \nii\ (22 regions) support this further, as the standard deviation about the gradient from these regions is $\sim$0.05 dex (see \S4.1). The other CHAOS galaxies show similarly low dispersion, ranging from $\sim$0.04 dex to 0.10 dex \citep{roge2021}.

Recently, \citet{este2022} examined the abundances in the Galaxy as measured from a number of sources, including \hii regions from \citet{arel2020M,arel2021}, with the goal of assessing the large abundance dispersion that \citet{deci2021} measure in Galactic neutral clouds. The direct abundances in the \hii regions are calculated using \te\nii\ in the low-ionization zone and \te\oiii\ in the high-ionization zone, only applying the \citet{este2009} \te-\te\ relation when one of these temperatures is missing. From the 42 Galactic regions, the dispersion about the O/H gradient is $\sim$0.07 dex, which is similar to the dispersion in metallicity measured from B-type stars, classical Cepheids, and young clusters in the Galaxy \citep{este2022}, and significantly less than the dispersion measured in the neutral clouds. \citet{mend2022a} corroborate the low dispersion in the \citet{arel2020M,arel2021} \hii regions when considering non-zero temperature fluctuations and updated \hii region positions. In the spiral galaxy M101, \citet{kenn2003b} determine the abundance gradient from 20 \hii regions, the majority of which have direct \te\oiii\ and \te\siii\ measurements and nine have direct \te\nii. Citing the tight correlation they observe between \te\oiii\ and \te\siii, \citet{kenn2003b} use an average of the direct \te\oiii\ and the inferred \te\oiii\ from \te\siii\ to obtain the high-ionization zone temperature and they use the direct \te\nii\ in the low-ionization zone, when available. Although this sample consists of only 20 regions, there is a clear abundance gradient in M101 with very little dispersion about the gradient. The dispersion reported by CHAOS \citep[0.097 dex,][]{roge2021} is slightly larger than the gradient from \citet{kenn2003b} would indicate, but this could be due to the inclusion of \hii regions with a single direct \te.

Taken together, the magnitude of the intrinsic dispersion in O/H observed in local spiral galaxies would suggest the following: 1. Large abundance variations are not typically observed when considering the \hii regions with the most reliable abundance measurements; 2. The mixing mechanisms in these galaxies act to efficiently homogenize the ISM such that the dispersion is at the level of observational uncertainty; 3. The use of single-temperature \hii regions has the potential to inflate the measured intrinsic dispersion to unphysically large values, erroneously suggesting that local enrichment/pollution are dominant relative to the mixing mechanisms. As discussed in \S5.1, the use of a single \te\oiii\ in M33 is particularly problematic due to various contaminating sources, which might explain the abundance scatter reported in previous studies. That being said, any \hii region could have a temperature structure that deviates slightly from the predicted \te-\te\ relations and can result in large abundance variations \citep{arel2020}. Robust \te-\te\ relations that account for other properties of the region (hardness of the ionizing radiation, metallicity, etc.) may improve inferred temperature estimates, but the simple linear relations have the potential to inflate abundance dispersion when only one temperature is applied. Strong line abundance calibrators have been shown to produce lower abundance variations relative to the direct technique when a single direct temperature is used \citep[see][]{arel2016}. However, empirical and theoretical strong line calibrators can produce different abundances, gradients, and dispersions in the same galaxy \citep{KE08,mous2010}. Therefore, for the most accurate measure of the abundance dispersion in a galaxy it is imperative to have direct temperature measurements across multiple ionization zones in a significant number of \hii regions.

\section{Conclusions}

We report on the CHemical Abundances Of Spirals observations of the well-studied spiral galaxy M33. Of the nearly 100 \hii regions targeted, we detect significant temperature-sensitive auroral line emission from \oiii, \nii, or \siii\ in 65 regions, as well as emission from the auroral lines of \oii, \sii, and [\ion{Ar}{3}]. We also detect [\ion{Cl}{3}]$\lambda\lambda$5517,5537 in 20 regions, allowing for the first statistically significant Cl abundance determination in a CHAOS galaxy. [\ion{Fe}{2}]$\lambda$4288 is significantly detected in 4 \hii regions, indicating that [\ion{Fe}{2}]$\lambda$4360 contamination of \oiii$\lambda$4363 must be accounted for in these regions, but that the degree [\ion{Fe}{2}] contamination is, in general, fairly minimal in M33.

The temperatures we measure in M33 show little scatter and generally follow the existing literature \te-\te\ relations. We use the direct temperatures from \oiii, \nii, and \siii\ and the \te-\te\ relations and weighted-average ionization zone temperature approach from \citet{roge2021}. The ionization zone temperatures are used to obtain the ionic abundances of many species in M33; the oxygen abundances measured in 65 \hii regions constitute the largest homogeneous sample of direct abundances in M33 to date.

From multiple elements, the chemical enrichment of M33 is homogeneous and shows little dispersion, indicating a well-mixed ISM. The intrinsic dispersion measured about the oxygen abundance gradient is consistent with the uncertainty on the most well-measured O/H. The scatter observed in the subsample of regions with direct temperatures of \te\oiii, \te\siii, and \te\nii\ is 0.05 dex, while the scatter measured in regions with fewer direct \te\ is closer to $\sim$0.11 dex. The secondary N/O gradient in M33 is shallower than the universal secondary N/O gradient reported in \citet{berg2020}, but is consistent with the gradient measured in other spiral galaxies with low stellar mass. The $\alpha$-process and $\alpha$-process-dependent elements show averages consistent with the solar values and do not vary as a function of galactocentric radius.

We compare our results to literature studies that measure oxygen abundance gradients in M33. While the total literature abundance sample is one of the largest in any spiral galaxy, it is obtained in an inhomogeneous manner (i.e., different detectors, wavelength coverage, and temperatures used) and a complete and consistent recalculation of temperatures and abundances is required to make a one-to-one comparison. We use the reported line intensities and complete this re-reduction, then compare the resulting temperatures and abundances to the CHAOS-measured values. While there is generally good agreement between the gradients determined from primarily direct-abundance studies, the direct abundance gradients determined from the primarily strong line studies are significantly shallower and have much larger dispersions.

Comparing the physical properties in common \hii regions, we find very good agreement with the \oiii\ and \nii\ temperatures and O/H abundances measured by \citetalias{tori2016}. The differences in O/H relative to the other literature studies are related, primarily, to the difference in \te\oiii. We discuss different mechanisms for \oiii$\lambda$4363 contamination, such as from [\ion{Fe}{2}]$\lambda$4360 or the \ion{Hg}{1} $\lambda$4358 sky line, which is a unique source of contamination for M33 due to its redshift. The use of low-resolution spectrographs can blend \oiii$\lambda$4363 with these contaminating sources and produce unphysical \te\oiii; using relatively high-resolution spectroscopy with broad wavelength coverage, sufficient to obtain electron temperatures from multiple ionization zones, is the only way to avoid, or limit the impact of, contamination.

The result that M33 is chemically well-mixed and shows little abundance variations is supported by the studies that meet the above criteria. Similar abundance studies that detect multiple temperatures from numerous \hii regions in local spiral galaxies also show intrinsic dispersions that are consistent with observational uncertainties. These findings suggest that the ISM is well-mixed in these galaxies and that the use of single-temperature regions can inflate the dispersion in O/H. Studies seeking to obtain the most accurate measure of the variation in O/H should strive to directly measure \te\ in multiple ionization zones to avoid any potential biases associated with the application of linear \te-\te\ relations or contamination in the single temperature.

We thank the referee for their thorough review of the original manuscript and the useful feedback that has improved the clarity of this work.
This work has been supported by NSF Grants AST-1109066 and AST-1714204. This paper uses data taken with the MODS built with funding from NSF grant AST-9987045 and the NSF Telescope System Instrumentation Program (TSIP), with additional funds from the Ohio Board of Regents and the Ohio State University Office of Research. This paper made use of the modsIDL spectral data reduction pipeline developed in part with 
funds provided by NSF Grant AST-1108693 and a generous gift from OSU Astronomy alumnus David G.
Price through the Price Fellowship in Astronomical Instrumentation. This work was based in part on observations made with the Large Binocular Telescope (LBT). The LBT is an international collaboration among institutions in the United States, Italy and Germany. The LBT Corporation partners are: the University of Arizona on behalf of the Arizona university system; the Istituto Nazionale di Astrofisica, Italy; the LBT Beteiligungsgesellschaft, Germany, representing the Max Planck Society, the Astrophysical Institute Potsdam, and Heidelberg University; the Ohio State University; and the Research Corporation, on behalf of the University of Notre Dame, the University of Minnesota, and the University of Virginia. 
This research has made use of the NASA/IPAC Extragalactic Database (NED) which is operated by the Jet Propulsion Laboratory, California Institute of Technology, under contract with the National Aeronautics and Space Administration.
This study has made use of the GTC Archive; the GTC Archive is part of the Spanish Virtual Observatory project funded by MCIN/AEI/10.13039/501100011033 through grant PID2020-112949GB-I00
\newpage
\appendix

In this Appendix, we provide the following information: Emission Line Detections are reported in Table \ref{t:m33Detect}; Line Intensities are provided in Table \ref{t:m33Int}; and the Temperatures and Abundances in each region are listed in Table \ref{t:m33Abun}.
%

\renewcommand{\thetable}{A.1}
\startlongtable
\begin{deluxetable*}{lccccccccccc}
\tablecaption{M33 Emission Line Detections}
\tablewidth{\textwidth}
\renewcommand{\arraystretch}{0.7}
\tablehead{
   \colhead{\hii}  & 
   \multicolumn{6}{c}{Auroral Lines}  & 
   \multicolumn{4}{c}{Other Lines}  & 
   \colhead{Wolf} \\\cmidrule(lr){2-7}\cmidrule(lr){8-11}
   \colhead{Region}  & 
   \colhead{\oiii}  & 
   \colhead{\nii}  & 
   \colhead{\siii}  & 
   \colhead{\oii}  & 
   \colhead{\sii}  & 
   \colhead{[\ion{Ar}{3}]}  & 
   \colhead{\ion{C}{2}}  & 
   \colhead{[\ion{Cl}{3}]}  & 
   \colhead{[\ion{Ar}{4}]}  & 
   \colhead{[\ion{Fe}{2}]}  & 
   \colhead{Rayet}}
\startdata
Total Detections:   & 28  & 42  & 62  & 84  & 64  & 5  & 10  & 20  & 4  & 4  &   \\ \hline
M33$-$2+25  &    &    &    & \checkmark  &    &    &    &    &    &    &   \\
M33+23$-$9  &    & \checkmark  & \checkmark  &    & \checkmark  &    &    &    &    &    & \checkmark  \\
M33$-$23+16  &    &    &    &    &    &    &    &    &    &    &   \\
M33$-$28+6  & \checkmark  & \checkmark  & \checkmark  & \checkmark  & \checkmark  &    &    & \checkmark   &    &    &   \\
M33$-$36$-$52  &    & \checkmark  & \checkmark  & \checkmark  & \checkmark  &    &    &    &    &    &   \\
M33+2+111  &    &    & \checkmark  & \checkmark  &    &    &    &    &    &    &   \\
M33$-$6$-$120  &    &    & \checkmark  & \checkmark  &    &    &    &    &    &    &   \\
\enddata
\label{t:m33Detect}
\tablecomments{Emission line detections in the \hii regions of M33. A checkmark indicates an emission line that is measured at S/N $>$ 3. Column 1: \hii region ID. Columns 2-7: Temperature-sensitive auroral line detections in the \hii region. The auroral lines are \oiii$\lambda$4363 (Col. 2), \nii$\lambda$5755 (Col. 3), \siii$\lambda$6312 (Col. 4), \oii$\lambda\lambda$7320,7330 (Col. 5), \sii$\lambda\lambda$4069,4076 (Col. 6), and [\ion{Ar}{3}]$\lambda$5192 (Col. 7). Columns 8-11: Significant detections of other physically-significant emission lines which include: \ion{C}{2} $\lambda$4267 (Col. 8), [\ion{Cl}{3}]$\lambda\lambda$5517,5537 (Col. 9), [\ion{Ar}{4}]$\lambda$4740 (Col. 10), and [\ion{Fe}{2}]$\lambda$4288 (Col. 11). Column 12: Wolf-Rayet features observed in the region.}
\end{deluxetable*}

\renewcommand{\thetable}{A.2}
\clearpage
\startlongtable
\begin{deluxetable*}{lccccccccc}
\tablewidth{\textwidth}
\tabletypesize{\footnotesize}
\tablecaption{M33 Emission Line Intensities}
\tablehead{
\vspace{-1ex} \\
\multicolumn{10}{c}{$I(\lambda)/I(\mbox{H}\beta)$}}
\setlength{\tabcolsep}{0.06pt}
\renewcommand{\arraystretch}{0.7}
\startdata
{Ion}               & M33$-$2+25          & M33+23$-$9          & M33$-$23+16          & M33$-$28+6          & M33$-$36$-$52          & M33+2+111          & M33$-$6$-$120          & M33+78+91          & M33$-$89$-$21         \\
\hline
{[O~\ii]~\W3727}    &  2.799$\pm$0.180          &  1.389$\pm$0.089          &  1.906$\pm$0.142          &  2.138$\pm$0.151          &  1.989$\pm$0.140          &  2.172$\pm$0.059          &  3.217$\pm$0.296          &  2.161$\pm$0.055          &  2.444$\pm$0.139         \\
{H13~\W3734}        &  0.027$\pm$0.004          &  0.027$\pm$0.006          &  0.030$\pm$0.008          &  0.030$\pm$0.003          &  0.026$\pm$0.003          &  29.4$\pm$1.8$\times$10$^{-3}$     &  0.028$\pm$0.006          &  27.0$\pm$2.0$\times$10$^{-3}$     &  0.030$\pm$0.002         \\
{H12~\W3750}        &  0.030$\pm$0.011          &  0.047$\pm$0.006          &  0.053$\pm$0.017          &  0.039$\pm$0.004          &  0.036$\pm$0.003          &  0.044$\pm$0.004          &  0.028$\pm$0.017          &  0.033$\pm$0.008          &  0.057$\pm$0.008         \\
{H11~\W3770}        &  0.046$\pm$0.009          &  0.058$\pm$0.004          &  0.056$\pm$0.013          &  0.047$\pm$0.004          &  0.043$\pm$0.004          &  0.047$\pm$0.005          &  0.036$\pm$0.012          &  0.047$\pm$0.004          &  0.060$\pm$0.006         \\
{H10~\W3797}        &  0.060$\pm$0.008          &  0.060$\pm$0.013          &  0.066$\pm$0.016          &  0.064$\pm$0.007          &  0.057$\pm$0.006          &  0.064$\pm$0.004          &  0.059$\pm$0.013          &  0.058$\pm$0.004          &  0.065$\pm$0.005         \\
{He~I~\W3819}       &  0.015$\pm$0.009          &  \nodata           &  0.013$\pm$0.015          &  0.017$\pm$0.004          &  0.011$\pm$0.003          &  0.011$\pm$0.003          &  0.017$\pm$0.006          &  0.013$\pm$0.003          &  0.002$\pm$0.002         \\
\enddata
\label{t:m33Int}
\tablecomments{The reddening-corrected emission line intensities measured in the \hii regions of M33. The left column provides the \hii region ID, the remaining columns provide the intensity for each emission line (rows). The last three rows provide the best-fit $c(H\beta)$, $a_H$ (the stellar absorption EW in \AA), and F$_{H\beta}$ in units of 10$^{-16}$ ergs s$^{-1}$ cm$^{-2}$.}
\end{deluxetable*}

\renewcommand{\thetable}{A.3}
\startlongtable
\begin{deluxetable*}{lccccccccc}
\tablewidth{\textwidth}
\tabletypesize{\footnotesize}
\tablecaption{M33 Temperatures, Densities, and Abundances}
\tablehead{\multicolumn{9}{c}{M33 Physical Properties}}
\setlength{\tabcolsep}{1pt}
\renewcommand{\arraystretch}{0.7}
\startdata
Property                    & M33+23$-$9          & M33$-$28+6          & M33$-$36$-$52          & M33+2+111          & M33$-$6$-$120          & M33+78+91          & M33$-$66$-$161          & M33+89+165         \\
\hline
$T_e$[S\ii]~(K)              &  5400$\pm$400        &  9300$\pm$1900        &  9200$\pm$900        &  \nodata              &  \nodata              &  11800$\pm$3400        &  8700$\pm$500        &  8700$\pm$400       \\
$T_e$[N\ii]~(K)              &  6700$\pm$400        &  8200$\pm$500        &  7800$\pm$500        &  \nodata              &  \nodata              &  9400$\pm$700        &  7900$\pm$200        &  8000$\pm$400       \\
$T_e$[O\ii]~(K)              &  \nodata              &  8900$\pm$400        &  8500$\pm$400        &  7200$\pm$300        &  7600$\pm$400        &  7700$\pm$300        &  8500$\pm$300        &  8200$\pm$200       \\
$T_e$[S\iii]~(K)             &  6200$\pm$600        &  8000$\pm$300        &  7000$\pm$200        &  7500$\pm$600        &  7400$\pm$400        &  6800$\pm$400        &  6900$\pm$100        &  7400$\pm$300       \\
T$_e$[O~\iii] (K)           &  \nodata              &  8100$\pm$700        &  \nodata              &  \nodata              &  \nodata              &  \nodata              &  \nodata              &  \nodata             \\
$T_e$[Ar\iii]~(K)             &  \nodata              &  \nodata              &  \nodata              &  \nodata              &  \nodata              &  \nodata              &  \nodata              &  \nodata             \\
$n_{e,[S II]}$~(cm$^{-3}$)        &  30$\pm$80        &  160$\pm$150        &  60$\pm$80        &  30$\pm$100        &  110$\pm$110        &  40$\pm$110        &  70$\pm$40        &  30$\pm$40       \\
\enddata
\label{t:m33Abun}
\tablecomments{The electron temperatures, electron densities, ionic abundances, and total abundances in the \hii regions of M33. The left column lists the property and the relevant units, the remaining columns provide the values of the properties for each region. The MCMC technique described in \S3.1 can result in large errors on $n_{e,[Cl III]}$; [\ion{Cl}{3}] densities with uncertainty larger than three times the calculated density are reported with ":".}
\end{deluxetable*}
\newpage
\bibliographystyle{aasjournal}
\bibliography{chaosvii_bib.bib}

\end{document}